\documentclass[thesis]{Classes/VTthesis}

\newcommand{\CommitteeHead}{Danfeng (Daphne) Yao}
\newcommand{\CommitteeMemberOne}{Na Meng}
\newcommand{\CommitteeMemberTwo}{Franciso J. Servant}
\newcommand{\Degree}{Masters of Science}
\newcommand{\Program}{Computer Science \& Application}
\newcommand{\ProjTitle}{Enhancing CryptoGuard's Deployability for Continuous Software Security Scanning}
\newcommand{\Self}{Miles Frantz}
\newcommand{\College}{Virginia Polytechnic Institute and State University}
\newcommand{\DefenseDate}{April 27, 2020}
\newcommand{\KeyWords}{Cryptoguard, Static-Code Analyzer, Java, Deployment Grade, Gradle, Maven, Java 8, Java 7, Java 11}

\newcommand{\Abbreviations}{
	\nomenclature{SDLC}{The natural \textbf{S}oftware \textbf{D}evelopment \textbf{L}ife\textbf{C}ycle, representing the stages of an application through planning, development, testing, review, and deployment.}

	\nomenclature{IDE}{An \textbf{I}ntegrated \textbf{D}evelopment \textbf{E}nvironment, which is a program that eases and enables people to write their programs or applications.}

	\nomenclature{Build Tool}{A program specially created for its intended programming language, to aide in various aspects throughout the SDLC.}

	\nomenclature{JDK}{The \textbf{J}ava \textbf{Development} \textbf{Kit}, which is used to write programs or applications in Java.}

	\nomenclature{JDK SE}{The JDK \textbf{S}tandard \textbf{E}dition, which usually used for non-web based applications.}

	\nomenclature{JDK EE}{The JDK \textbf{E}nterprise \textbf{E}dition, which is usually used for web-based applications.}

	\nomenclature{LTS}{\textbf{L}ong \textbf{T}erm \textbf{S}upport.}

	\nomenclature{JVM}{The \textbf{J}ava \textbf{V}irtual \textbf{M}achine, which is the program which executes Java programs.}

	\nomenclature{XML}{E\textbf{x}tensible \textbf{M}arkup \textbf{L}anguage, a highly common and configurable markup language, fundamentally used to help represent configuration and web pages.}

	\nomenclature{JSON}{Similar to XML is another markup language commonly used for configuration.}

	\nomenclature{Control Flow Graph}{A representation of the different paths that a program may take.}

	\nomenclature{Marshalling}{Converting raw information or data into a specific form.}

	\nomenclature{Unmarshalling}{Converting information from a specific form into raw objects.}

	\nomenclature{JEP}{The \textbf{J}DK \textbf{E}nhancement \textbf{P}roposal, a process to create an incremental process upon Java.}

	\nomenclature{SDK}{\textbf{S}oftware \textbf{D}evelopment \textbf{K}it, used to help create programs.}

	\nomenclature{Compile}{Creating an executable program from various source code.}

	\nomenclature{Decompile}{Taking an executable program and interpreting the source code that likely built it.}

	\nomenclature{Ant}{\textbf{A}nother \textbf{N}eat \textbf{T}ool, a build tool developed by Apache}

	\nomenclature{DoD}{\textbf{D}epartment \textbf{o}f \textbf{D}efense}

	\nomenclature{DHS}{\textbf{D}epartment of \textbf{H}omeland \textbf{S}ecurity}

	\nomenclature{VM}{\textbf{V}irtual \textbf{M}achine, a program that emulates a system and executes a certain set of instructions.}

	\nomenclature{JVM}{\textbf{J}ava \textbf{V}irtual \textbf{M}achine, the program that executes Java code.}

	\nomenclature{WebApp}{A program that operates through web based protocols or hosted on a browser.}

	\nomenclature{IoT}{\textbf{I}nternet \textbf{o}f \textbf{T}hings, low-power devices (like a raspberry pi) communicating accross internet protocols.}

}
\usepackage[export]{adjustbox}
\usepackage{listings}
\usepackage{float}
\usepackage{titlesec}

\titlespacing*{\section}{0pt}{0ex}{0ex}
\titlespacing*{\subsection}{0pt}{0ex}{-1ex}
\titlespacing*{\subsubsection}{0pt}{0ex}{-2ex}

\def\BibTeX{{\rm B\kern-.05em{\sc i\kern-.025em b}\kern-.08em
T\kern-.1667em\lower.7ex\hbox{E}\kern-.125emX}}

\definecolor{codegreen}{rgb}{0,0.6,0}
\definecolor{codegray}{rgb}{0.5,0.5,0.5}
\definecolor{codepurple}{rgb}{0.58,0,0.82}
\definecolor{backcolour}{rgb}{0.95,0.95,0.92}

\lstdefinestyle{mystyle}{
	basicstyle=\tiny,
	backgroundcolor=\color{backcolour},
	commentstyle=\color{codegreen},
	keywordstyle=\color{magenta},
	numberstyle=\tiny\color{codegray},
	stringstyle=\color{codepurple},
	breakatwhitespace=false,
	breaklines=true,
	captionpos=b,
	keepspaces=true,
	numbers=left,
	numbersep=3pt,
	showspaces=false,
	showstringspaces=false,
	showtabs=false,
	tabsize=1
}

\title{\ProjTitle}

\keywords{\KeyWords}

\author{\Self}

\program{\Program} 

\degree{\Degree} 

\submitdate{\DefenseDate} 

\institution{\College}

\principaladvisor{\CommitteeHead}
\firstreader{\CommitteeMemberOne}
\secondreader{\CommitteeMemberTwo}

\dedication{
	I dedicate this to my family and friends, both of which helped encourage me to follow my passion.
	I would also like to dedicate this to one of my mentors from an internship, who taught me the important lesson ``if something is not easy to use, it will not be used''.
}

\acknowledge{
	I would like to acknowledge Sazzadur Rahaman, since his mentoring helped guide me throughout the overall research process.
	Next, I would also like to acknowledge Dr. Daphne Yao, who has helped encourage me to overcome my weaknesses.
	Finally, I would like to acknowledge the Continuous Assurance team, whom have been and continue to have patience while working collaboratively.
}

\abstract{
	The increasing development speed via Agile\cite{Agile_Cycle} may introduce overlooked security steps in the process, with an example being the Iowa Caucus application\cite{Koebler2020a}.
	Verifying the protection of confidential information such as social security numbers requires security at all levels, providing protection through any connected applications.
	CryptoGuard\cite{Rahaman2019}\footnote{located at \href{https://github.com/CryptoGuardOSS/cryptoguard}{https://github.com/CryptoGuardOSS/cryptoguard}} is a static code analyzer for Java.
	This program verifies that developers do not leave vulnerabilities in their application.
	The program aids the developer by identifying cryptographic misuses such as hard-coded keys, weak program hashes, and using insecure protocols.
	In my Master thesis work, I made several important contributions to improving the deployability, accessibility, and usability of CryptoGuard.
	I extended CryptoGuard to scan source and compiled code, created live documentation, and supported a dual cloud and local tool-suite. I also created build tool plugins and a program aid for CryptoGuard.
	In addition, I also analyzed several Java-related surveys encompassing more than 50,000 developers and reported interesting current practices of real-world software developers.
} \label{abstract}

\abstractgenaud{
	Throughout the rise of software development, there has been an increase in development speed with developers embracing methodologies that use higher rates of changes \cite{Agile_Cycle}, such as Agile.
	Since Agile naturally addresses ``problems of rapid change'' \cite{Agile_People}, this also increases the likelihood of insecure and vulnerable coding practices.
	Though consumers depend on various public applications, there can still be failures throughout the development process in applications such as the Iowa caucus application.
	It was determined the Iowa cacus application development teams' repository credentials (API key) was left within the application itself \cite{Koebler2020a}.
	API keys provide the credential to be able to directly interact with server systems, and if left unguarded can be easily exploited.
	Since the Iowa cacus application was released publicly, malicious actors (other people looking to exploit the application) may have already discovered this credential.
	Within our team we have created CryptoGuard \cite{Rahaman2019}, a program to analyze applications to detect cryptographic issues such as an API key.
	Creating it with scalability in mind, it was created to be able to scan enterprise code at a reasonable speed.
	To ensure its use within companies, we have been working on extending and enhancing the work to the current needs of Java developers.
	Verifying the current Java landscape, we investigated three different companies and their developer ecosystem surveys that are publicly available.
	Amongst these companies are; JetBrains, known for their Integrated Development Environments (IDE, or application to help write applications) and their own programming language, Snyk, known for their public security platform and anti-virus capability, and Jakarta EE, which is the new platform for the enterprise version of Java.
	Throughout these surveys, we accumulate more than 50,000 developers' responses, spanning various countries, company experience, and ages.
	With their responses amalgamated, we enhance CryptoGuard to be available to as many developers and their requests as possible.
	First, CryptoGuard is enhanced to scan a projects source code.
	After that, ensuring our project is hosted by a cloud service, we actively are extending our project to the Security Assurance Marketplace (SWAMP).
	Funded by the DHS, SWAMP not only supplies a public cloud for developers to use, but a local download option to scan a program within the user's own computer.
	Next, we create a plugin for two most used build tools, Gradle and Maven.
	Then to ensure CryptoGuard can be have reactive aide, CryptoSoule is created to aide minimal interface aide.
	Finally utilizing a live documentation service, an open source documentation website was created to provide working examples to the community.
} \label{abstract_general}

\begin{document}
\frontmatter
\maketitle
\tableofcontents

\listoffigures

\printnomenclature 

\Abbreviations

\mainmatter

\chapter{Introduction} \label{ch:introduction}
\section {Introduction} \label{se:intro}
Despite all the publicly available information and automatic programming tools, there are still an abundance of vulnerabilities within publicly used programs.
Not only are these vulnerabilities limited to smaller developers, they are proliferating throughout companies potentially due to their size and the quicker development life cycles provided for product development teams.
With a limited scaling to smaller projects, many automatic programming tools that scan projects source for cryptographic misuse (static code analyzers) cannot handle the major scale of projects that companies produce.
To combat this, CryptoGuard was created (within the same lab) as a static code analyzer to not only more effectively but also more efficiently scan major project source.
Limited in its first release, it could only scan certain types of Java projects source using a strict input and output format, further narrowing its target audience.
Within this thesis work, we enhanced CryptoGuard's abilities to not only extend what it could scan alongside the developer, but also allow it to ascend to the cloud to be able to be publicly used by even more developers.

\section{Contributions} \label{se:contributions}
Our major contributions onto CryptoGuard are listed below\footnote{our enhancements can be found at \href{https://github.com/franceme/cryptoguard}{https://github.com/franceme/cryptoguard}}:

\begin{enumerate}
	\item We added the ability to scan the Java files and Java class files into the project.
	\item We created output formats to supply information to various consumers.
	\item We created a plugin for both Maven and Gradle to integrate CryptoGuard in each of the build tools.
	\item We create a Jupyter Notebook for live documentation.
	\item We create a public package release in GitHub Repository.
\end{enumerate}

\section{Thesis Layout} \label{se:Layout}

The rest of this thesis follows this chronological order.
First chapter \ref{ch:lit_review} goes over the core concept of this thesis and several other pertinent documents that supply insight towards our direction.
Next chapter \ref{ch:swamp} describes earlier, current, and ongoing work of enhancing CryptoGuard to be able to work on a publicly available cloud platform.
Then chapter \ref{ch:survey} analyzes the current Java landscape by investigating several public surveys individually and in conglomeration.
After that, chapter \ref{ch:enhancements} encompasses other enhancements put pulled into CryptoGuard to ensure developers have an easier method of both retrieving and using CryptoGuard within their development.
Finally, chapter \ref{ch:discussion} describes some of the key issues that arose during this enhancement process and describes any future work from this project.

\chapter{Review of Literature} \label{ch:lit_review}
\section{Detecting Vulnerabilities Within Java} \label{se:Intro_Vuln}
Majority of the vulnerabilities that seem appear in the news are due to simple issues.
These issues include using insecure internet protocols or weak passwords.
Throughout several sources, one of most recent examples of this is the massive Equifax breach \cite{Franceschi-Bicchierai2017}.
Still feeling the effects of the vulnerability, Equifax is one of the major Credit Companies within the United States and was the target of a major breach.
One security researcher was able to find a public website that gained access to ``the personal data of every American, including social security numbers, full names, birth dates, and city and state of residence.'' \cite{Franceschi-Bicchierai2017}.
The team at VICE \cite{Franceschi-Bicchierai2017} discovered the data included more than 145 million American records.
Since a company has that much pertinent and privileged information to lose, someone would expect their product and platform to have thorough security practices.
Not only was the breach possible due to an out of date software Apache package \cite{Franceschi-Bicchierai2017}, another website created to mitigate claims also used the basic credentials of ``admin'' \cite{Newman2017}.

A more recent and potentially less dangerous example is the recent Iowa Caucus application created for the 2020 Iowa Democratic Caucus.
Shadow Inc. \cite{Koebler2020} created this application to not only generate the results quickly, the company also promised the results would be secure and verified.
Ideally this application counts the votes correctly as well as returning the results promptly.
Unfortunately, the development team rushed the application and the team at Vice \cite{Koebler2020a} described the article as an ``off the shelf skeleton project app''.
Not only was it rushed, the technological skill appraisal does not inspire such confidence.
With an intense sense of dejection, the team at Vice \cite{Koebler2020a} further described it as being ``a starter package and they just added things on top of it''.
None of this information determines the application vulnerable to any attacks yet, proves the effort put into development.
Given the low effort put into developing the application, necessities are known to slip and forgotten about to ensure a working product.
Described as correctly configured, found within the application itself is the hard-coded API key.
Not necessarily directly creating a vulnerability, the researchers found ``potentially concerning code within it, including hard-coded API keys'' \cite{Koebler2020a}.
Since this is a potential vulnerability, other researchers, penetration testers, or more malicious attackers may have discovered it already.
A lack of specific domain knowledge or using less security focused advice from a public forum such as Stack Overflow may attributed to such vulnerabilities.
Despite creating a corrective machine learning algorithm to suggest more secure recommendations only 10.1 percent or approximately 3 participants used the more secure recommendations \cite{Meraki2019}, while the others used known insecure responses.
Having a tool to automatically detect these issues instead of relying on external sources would have alerted the developers to the issue before releasing it.

\section{Analysis Tools} \label{se:Intro_CodeAnalyzers}
Amongst various popular tools that developers use to automatically detect security issues, static code analysis and dynamic code analysis are the two prevalent methodologies \cite{Rahaman2019}.
Each method has its specific use case and tailored for a specific scenario.
Static code analysis requires the program to scan each file.
Scanning through source files enables a tool to handle enterprise project scale projects while minimizing any missed (false negative) security issues.
Scaling to massive project sizes also takes more time, which is avoided by dynamic code analysis.
Dynamic code analysis does not require the program, instead ``watching'' a programs execution by various means\footnote{Trailing logs, call stack, VM logs, etc. ...} and when the program calls the specified method scanning the program and execution.
Supplying a ``reactive'' approach to scanning, minimize fake issues (false positive) \cite{Rahaman2019}.
Based on the specification of methods to wait for, dynamic code analysis tools are not as exhaustive as static code analysis tools.

Several teams have discovered many developers do not use static code analysis tools for assorted reasons despite the availability and benefits of static code analysis tools.
Developers are resistant to using such tools for multiple reasons security scanning tools, including each analysis tool will add time onto the compilation or integration testing phase.
Since static code analysis tools may provide many warnings or false positives, developers have stated being less averse to usage if the output displayed is more user-friendly \cite{10.5555/2486788.2486877}.
A more obtuse output would add more effort onto the developer and take more time away solely to understand the results.
This added time would also extend the software development life cycle (SDLC) and disrupt the developers' workflow, as acknowledged by majority of the developers from \cite{10.5555/2486788.2486877}.
A cryptographic tool would be able to address these problems and avoid such common pitfalls.

\section{CryptoGuard} \label{se:Intro_Crypto}
There is a various multitude of source-code analysis tools, regardless of the language or platform.
For example, Open Web Application Security Project (OWASP) \cite{OWASP} a useful and prevalent platform for helping encourage and teach security practices.
Listed on the website, there are 55 \cite{OWASPTool} active Static Code Analysis Tools, of which only 43 percent are open source tools and of those only 45 percent support the Java language.
Combining these two, only about 20 percent of the tools are both Open Source and support the Java language.

An upcoming cryptographic static code analyzer, CryptoGuard is a project created from Sazzadur Rahaman \cite{Rahaman2019} as a static code analyzer that scans projects against 16 specific rules \cite{Rahaman2019}.
Throughout the original study some of the more popular static code analysis tools (such as CrySL, and CogniCrypt \cite{Rahaman2019}) are not optimized for large code bases.
Progressing on top of this problem, CryptoGuard was able to successfully scan 46 Apache projects and 6,181 Android apps \cite{Rahaman2019a}.
CryptoGuard scanned several enterprise level Apache projects in an average of 3.3 minutes \cite{Rahaman2019a}.
CryptoGuard also scanned several android projects (automatically exiting after 10 minutes) with a runtime of 3.2 minutes \cite{Rahaman2019a}.
These results were achieved by using a stricter control-flow graph, only specifying certain security related APIs.
With a stricter and therefor smaller control-flow graph, the program has less information to slice through to find the vulnerabilities, making it both more precise and accuracy.
CryptoGuard is limited to scanning Java projects version 8 and below due to it is internal libraries.
Fortunately, Java 8 is a long-term support version and is theorized to be the most used version.
This theory is confirmed through several surveys in chapter \ref{ch:survey}.
Enough for the first comparisons CryptoGuard was able to scan Android Apps, compiled Jar files, and project directories through the command line.
Limiting the project to the command line directly lowers its developer outreach.
Throughout the rest of this section we will explain the groundwork on how we will expand the accessibility of the project.

\section{Java Landscape} \label{se:Intro_JavaDomain}
\vspace*{15pt}\begin{figure}[H]
	\begin{center}
		\includegraphics[width=\textwidth]{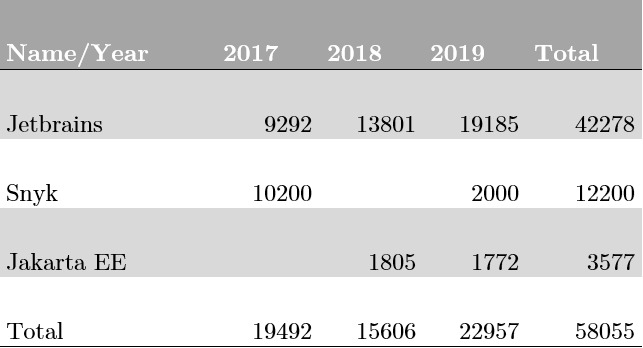}
	\end{center}
	\caption{Each of the surveys distributors and their participants.}
	\label{fig:DistributorsSurvey0}
\end{figure}
Created from around 1995 \cite{Oraclea}, Java has had a striving community surrounding it is creating libraries and build tools to help the community.
Though initially growing slowly, its adoption through Java Applets helped to spread its technological adoption and was a common build tool throughout many operating systems.
Throughout Javas history and development, there are more than 12 million developers using Java and more than 5 million students studying Java \cite{Oracle}.
Throughout all these users under the following section \ref{se:SurveyGeneral}, delving into a collective survey of at least 50,000 participants across a time span of 3 years.
Though we did not distribute the surveys ourselves, the statistics taken and studied are common throughout majority of the surveys and do not use distinct products of the distributor.
Listed in figure \ref{fig:DistributorsSurvey0} are the distributors listed and the number of participants per year.

\section{Java Build Tools} \label{se:Intro_Build}

\subsection{Which Build Tool to Use?}
Build tools are programs created to assist with the SDLC of a project in various ways.
For example, some of the more popular and recent build tools can provide the following benefits:

\begin{enumerate}
	\item Compile the project (with or without the use of a configuration file) to a specific architecture.
	\item Automatically run all the project tests associated with a project.
	\item Deploy the project (including any associated files) to a centralized (or specified) repository.
	\item Download outlined project dependencies from a centralized (or specified) repository.
	\item Format the project files to a specific style.
\end{enumerate}

With the backing of Javas community, there have been several different build tools each with their unique features.
Listed below are some of the main build tools that have been created for the Java Virtual Machine (JVM) based languages.

\subsubsection{Another Build Tool (Ant)}
One of the original build tools created for Java projects, Ant was initially released in 2000 by the Apache project.
The native functionality provided by Ant is like a configuration file, as it used an XML to specify the compilation arguments and the location of dependencies.
This was created as an improvement onto the already existing Make and targeted for the JVM family.

\subsubsection{Maven}
The second and one of the more popular build tools, Maven was created in 2002 \cite{Javatpoint} by the Apache project \cite{Apache} as well.
This tool uses an XML file for configuration as well, however it also has a centralized repository for dependencies, so it provides the ability for relatively easy dependency management.
The dependency management also provides hashing of the dependencies to ensure they are safe and have been verified.
Its framework also provides build commands to automatically run the tests listed within the project and can easily handle community made plugins.

\subsubsection{Gradle}
The third and relatively second most popular build tool for Java, Gradle \cite{GradleHome} was created in 2012 \cite{Gradle2020} by the Gradle Company with support of another programming language backing it.
Instead of using a standard XML file configuration, Gradle uses the programming language Groovy (created by Apache) to read a custom configuration file.
Gradle provides the same kind of build features (such as dependency management and running tests) as well as letting the user more easily create custom rules within the configuration file itself.
Any custom rule (or macro)  created is run as valid Groovy code.

\subsubsection{Scala Build Tool (SBT)}
Another build tool associated with Java is SBT, which was created in 2017 \cite{LightbendRelease} by Lightbend Inc \cite{Lightbend}.
Though this tool was primarily created for the Scala programming language (a Java-based programming language), this has also been used for Java projects as well.
This tool provides the same kind of features that Gradle provides, though with the backing of the Scala programming language instead of Groovy.

\subsubsection{Bazel}
Though the relative underdog, Bazel has been picking up a dedicated community since its creation in late 2019 \cite{GoogleBazelRelease} by an open source group initially started by Google \cite{GoogleBazel}.
Though majority of the code is open source, unlike others on the list parts of the repository are still stated to be closed source (due to licensing issues with Google).
Though being the \textit{youngest} build tool amongst those studied, appearing on the latest survey with a group dedicated group is promising.
Focusing on scalability and build time, it promises to build only portions of the code that have changed and be able to scale to any enterprise mono-repository size.

\chapter{Software Assurance Marketplace} \label{ch:swamp}
\section{What is Software Assurance Marketplace (SWAMP)?} \label{se:SWAMP}
Further extending the programs availability, there is ongoing progress to put this project into the popular provider \textbf{S}oft\textbf{w}are \textbf{A}ssurance \textbf{M}arket\textbf{p}lace \cite{SWAMP}.
Originating from and originally funded from the Department of Homeland Security (DHS) \cite{DHS2011}, this program was created via security specialists to provide an exhaustive and comprehensive security service to ensure the security of programs.
Not only does the project meet the major requirements for handling homeland security projects life cycle (research, development, test, and evaluation), SWAMP is an open source platform allowing for developers to both use it for testing their source code and for security developers to create and deploy their own tools.
With the cloud platform (\href{https://www.mir-swamp.org/}{MIR Swamp} \cite{MIRSwamp}) created early in February 2016 \cite{SWAMP}, the team also created a smaller yet locally hosting platform \href{https://continuousassurance.org/swamp-in-a-box/}{SWAMP-in-a-box} in late 2016 \cite{SWAMP}.
Creating the \href{https://continuousassurance.org/swamp-in-a-box/}{SWAMP-in-a-box} provides a more isolated and secure service that allows tools to be scanned privately, in a more secure environment.
Envisioned to host a wide set of publicly available tools, there are currently more than 500 tools available to be used in both cloud and non-cloud or localized (SWAMP-in-a-box) environments.
Popular tools such as Findbugs and Spotbugs are hosted here, as well as other tools that span a multitude of languages (not solely the Java programming language).
Utilizing this kind of platform, CryptoGuard would be able to greatly increase it is impact, as will be stated later the working environment is not obvious.

\section{Scanning Enhancements} \label{se:ScanningFiles}

Initially at its paper submission, CryptoGuard supported scanning Android APKs, Java Jars, and Maven or Gradle Projects.
While following a strict argument pattern, the executable provides enough arguments to scan the three available project sources.

\vspace*{15pt}\begin{figure}[H]
	\begin{center}
		\includegraphics[width=\linewidth]{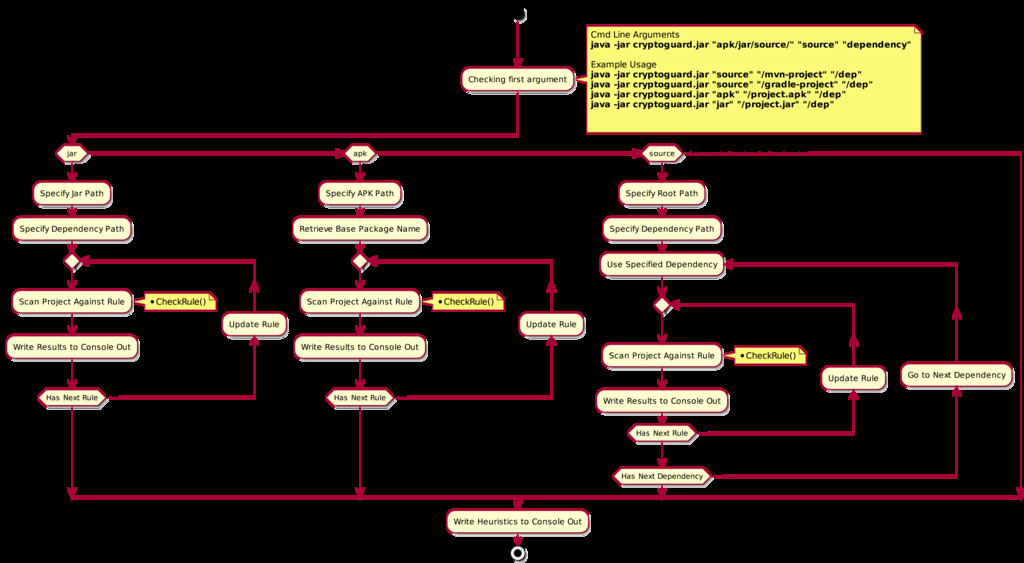}
	\end{center}
	\caption{The general flow of the original design.}
	\label{fig:Rigorityj}
\end{figure}

\begin{figure}[H]
	\begin{center}
		\includegraphics[width=\linewidth]{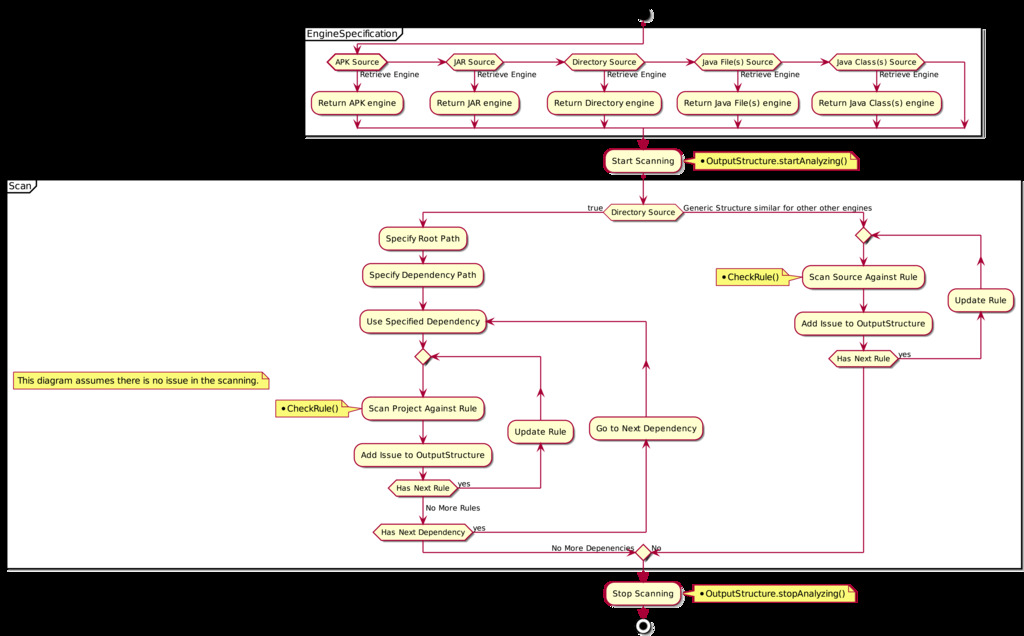}
	\end{center}
	\caption{The expansion of source scanning types.}
	\label{fig:EngineSpec}
\end{figure}

As shown in the activity diagram \ref{fig:Rigorityj}, the interface for the project is explicit and strict.
Only supplied are the three paths that can be utilized, which are expanded upon and explored more in the following sections.
Shown below (from figure \ref{fig:EngineSpec}) is the excerpt from the overall diagram that is specifically pertinent this section.

\vspace*{15pt}

To stand upon the shoulders of giants, CryptoGuard depends on the Soot \cite{McGillUniversity/SableResearchGroup} library to build the control flow graph.
Actively maintained throughout their \href{https://github.com/Sable/soot}{GitHub}, Soot (being used by the command line) takes a compiled Java Program and returns one of four different representations for the control flow graph of a program.
Like a few of the other programs using the project, CryptoGuard extends it and actively uses specific public methods provided within the library.
While directly extending Soot instead of calling it directly from the front end (command line), artifacts or compiled java code must be tacked into the Soot class path.
Any files along the Soot class path are directly loaded into memory and are therefore able to be read by Soot.
In addition, the class path requires the fully qualified names, which represents the package name or the sub-folder that contains the java class in relative to the base project directory.
Due to this tightly coupled relationship CryptoGuard is limited to any deficiencies that Soot has, including the supporting Java Version.
Remembering the traction created from Java Versions in section \ref{se:Intro_JavaDomain}, it can be hard for smaller projects and teams to ensure they are compatible with the newer Java Versions.
Still affected by this transition and the Jigsaw change since the update in Java 9, at the current time of this writing Soot is still in development to include the features of the Jigsaw features.
Unfortunately, since Java is currently on the \href{https://openjdk.java.net/projects/jdk/14}{non-LTS Version 14}, this means Soot is potentially 6 major versions and 1 major LTS out of date.
This is not a showstopper as summarized later in section \ref{se:SurveyGeneral}, as Java 8 is still the Java Version used by solid majority.
Since this means Soot still has a major impact upon the development community for now, CryptoGuard's use of Soot will be used more to reach more developers.

\subsection{Java Files} \label{sse:JavaFiles}

\vspace*{15pt}\begin{figure}[H]
	\begin{center}
		\includegraphics[scale=.75]{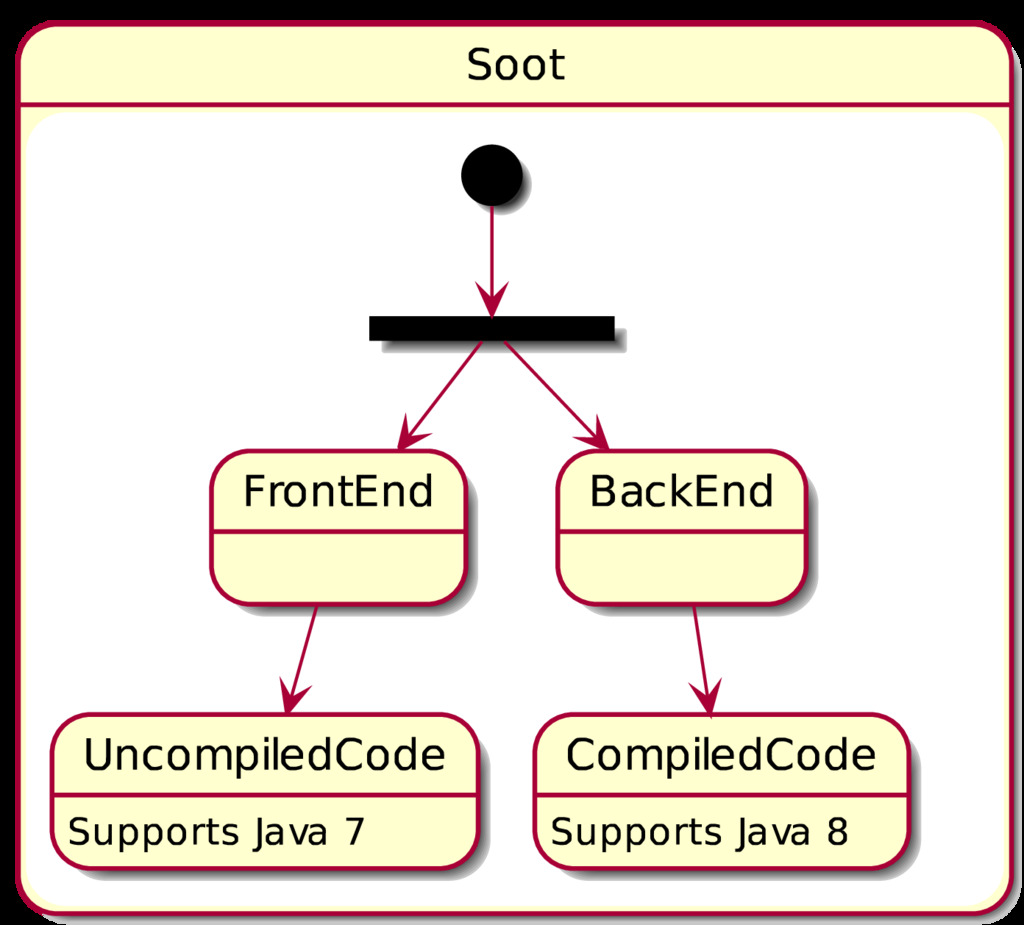}
	\end{center}
	\caption{Limited usages within the Soot library.}
	\label{fig:SootEngine}
\end{figure}

One of the main avenues to allow better use cases for SWAMP is to enable scanning raw Java Files.
Within their inquiry about the abilities of CryptoGuard, this was one of the initial direct requests.
Within Soot and due to the number of JEPs that have been introduced, it has taken all the teams' time to work diligently upon the back end.
Depicted within figure \ref{fig:SootEngine} is the program support provided within the current state of Soot.
This was confirmed \cite{GithubFrantz2019} within their active GitHub account and within our early on collaboration to determine the usages we could pull into CryptoGuard.
Within the self-provided issue were several other comments spanning similar issues encompassing the same misunderstanding-understanding of the front end of the project.

\vspace*{15pt}\begin{figure}[H]
\begin{lstlisting}[language=java, linerange={263-263, 274-296}]
  public static void setupBaseJava(
    Options.v().set_src_prec(Options.src_prec_java);
    Options.v().set_output_format(Options.output_format_jimple);

    Options.v().set_verbose(true);
    Options.v().set_validate(true);
    Options.v().set_whole_program(true);

    List<String> classNames = Utils.retrieveFullyQualifiedName(snippetPath);

    Scene.v()
        .setSootClassPath(
            Utils.surround(
                ":",
                Utils.getBaseSoot(javaHome),
                Utils.joinSpecialSootClassPath(snippetPath),
                Utils.buildSootClassPath(projectDependency)));
    log.debug("Setting the soot class path as: " + Scene.v().getSootClassPath());

    for (String dependency : Utils.getJarsInDirectories(projectDependency)) {
      classNames.addAll(Utils.getClassNamesFromJarArchive(dependency));
    }

    loadBaseSootInfo(classNames, criteriaClass, criteriaMethod, criteriaParam, checker, mainKlass);
\end{lstlisting}
	\caption{The method to setup the Soot environment for Java files.}
	\label{fig:BaseRouter_Java}
\end{figure}

While the back end of the project supplied enough exposure for Java Projects, this was similar enough mimic functionality for Java Files.
Within the code excerpt of \ref{fig:BaseRouter_Java} is the manner of manually setting up the proper Soot environment variables.
Separating itself from the Java Project is the manual composition of the Soot class path, in line \#11 to line \#22.
Explicitly naming the fully qualified names in the Soot class path is necessary to avoid scanning any other files within the same directory (or nested directories).
This also enables these classes to be picked up within Soot via the code interpreter and to be decompiled.

\vspace*{15pt}\begin{figure}[H]
\begin{lstlisting}[language=java, firstline=651, lastline=672]
    if (in.toLowerCase().endsWith(".java")) {
      sourcePackage = sourcePackage.replace(".java", "");
      try (BufferedReader br = new BufferedReader(new FileReader(in))) {

        String firstLine = br.readLine();
        Matcher match = null;
        while (StringUtils.isBlank(firstLine)
            || (match = startComment.matcher(firstLine)).find()
            || (match = comment.matcher(firstLine)).find()) firstLine = br.readLine();

        if (firstLine.startsWith("package ") && firstLine.toLowerCase().endsWith(";")) {
          sourcePackage =
              firstLine.substring("package ".length(), firstLine.length() - 1)
                  + "."
                  + sourcePackage;
        }

      } catch (IOException e) {
        log.fatal("Error parsing file: " + in);
        throw new ExceptionHandler("Error parsing file: " + in, ExceptionId.FILE_READ);
      }
    } else if (in.toLowerCase().endsWith(".class")) {
\end{lstlisting}
	\caption{Retrieving the fully qualified path for Java files.}
	\label{fig:Utils_FullyQual_Java}
\end{figure}

The manner of retrieving the fully names are shown (for a single file) are shown in figure \ref{fig:Utils_FullyQual_Java} for Java Files.
Using a defensive strategy with the file checks, the java file is loaded into memory and manually parsed to retrieve the package name.
While line \#10 ensures any documentation or white space is ignored and passed, line \#13 verifies the existence of a package declaration before retrieving the package name space.
If the check fails, then it immediately stops reading the file in a fail fast manner.

\subsection{Java Class Files} \label{sse:JavaClassFiles}

Not a requested item from SWAMP but an iterative step in the process, scanning Java Class files was the next inevitable extension.
Being able to scan Java 8 JEPs (depicted within figure \ref{fig:SootEngine}), Java Class files not only opens the avenues for SWAMP, it also extends possibilities later described within the paper in section \ref{se:Plugins}.
Following a similar pattern to the format of extending CryptoGuard for Java files subsection \ref{sse:JavaFiles}, creating the extension for Java Class files followed the methodology for scanning Java Jars.

\vspace*{15pt}\begin{figure}[H]
\begin{lstlisting}[language=java, linerange={316-316, 327-353}]
  public static void setupBaseJavaClass(
    Options.v().set_src_prec(Options.src_prec_only_class);
    Options.v().set_output_format(Options.output_format_jimple);

    Options.v().set_verbose(true);
    Options.v().set_validate(true);
    Options.v().set_whole_program(true);

    List<String> classNames = Utils.retrieveFullyQualifiedName(sourceJavaClasses);

    Scene.v()
        .setSootClassPath(
            Utils.surround(
                ":",
                Utils.joinSpecialSootClassPath(sourceJavaClasses),
                Utils.getBaseSoot(javaHome),
                Utils.join(":", Utils.getJarsInDirectories(projectDependencyPath))));
    log.debug("Setting the soot class path as: " + Scene.v().getSootClassPath());

    for (String clazz : classNames) {
      log.debug("Working with the full class path: " + clazz);
      Options.v().classes().add(clazz);
    }

    for (String dependency : Utils.getJarsInDirectories(projectDependencyPath))
      classNames.addAll(Utils.getClassNamesFromJarArchive(dependency));

    loadBaseSootInfo(classNames, criteriaClass, criteriaMethod, criteriaParam, checker, mainKlass);
\end{lstlisting}
	\caption{The method to setup the Soot environment for Java class files.}
	\label{fig:BaseRouter_Class}
\end{figure}

From within the explicit code listing in figure \ref{fig:BaseRouter_Class}, the two major portions enabling this additional enhancement are listed at lines \#15 and \#22.
In opposition to the Java Jar class enumeration, the java class fully qualified class name must be explicitly retrieved than using Soot to read the manifest of the Java Jar file.
Once these fully qualified class names are retrieved and combined, they are explicitly added onto the Soot class path, in a similar manner for Java Files section \ref{sse:JavaFiles}.
Like Java Files as well, they are explicitly added to avoid scanning other files from within the directory.

\vspace*{15pt}\begin{figure}[H]
\begin{lstlisting}[language=java, firstline=672, lastline=683]
    } else if (in.toLowerCase().endsWith(".class")) {
      try {
        ClassReader reader = new ClassReader(new FileInputStream(in));
        sourcePackage = reader.getClassName().replace(fileSep, ".");
      } catch (FileNotFoundException e) {
        log.fatal("File was not found " + in);
        throw new ExceptionHandler("File " + in + " not available.", ExceptionId.FILE_READ);
      } catch (IOException e) {
        log.fatal("Error parsing file: " + in);
        throw new ExceptionHandler("Error parsing file: " + in, ExceptionId.FILE_READ);
      }
    }
\end{lstlisting}
	\caption{Retrieving the fully qualified path for Java class files.}
	\label{fig:Utils_FullyQual_Class}
\end{figure}

Shown above in the code listing \ref{fig:Utils_FullyQual_Class} is the retrieval of the fully qualified path of a compiled java class.
The previous method of dynamically determining the fully qualified path was replaced by decompiliation.
Decompiliation is the act of re\-creating the source from a binary file.
Decompiling the binary java class file provides a better solution than the fully qualified path retrieval based on the folder structure.
Using the general directory setup of the project is prone to incidents.
When users have a custom project setup, the fully qualified path retrieval is likely to fail.
This is all rectified by the files decompiliation.

\subsection{Schema Design} \label{sse:schema}
To ensure developers can create tools to provide information into SWAMP, it has an XML schema base standard that scan tools must conform to in order to be interpreted in the platform.
Fortunately, SWAMP has created various libraries for reading and writing these formats and have published these open standards on GitHub \cite{ContinuousAssuranced}.
Located at their \href{https://github.com/mirswamp/resultparser/blob/master/xsd/scarf\_v1.2.xsd}{GitHub Repository} is the latest schema that is used for interacting with the swamp system.
Early within our collaboration and integration, there were a few additional fields they required, and we provided, which necessitated the need for a minor version increase (from v1.1 to v1.2).

\subsection{Output Structure} \label{sse:output}
\begin{figure}[H]
	\begin{center}
		\includegraphics[width=\linewidth]{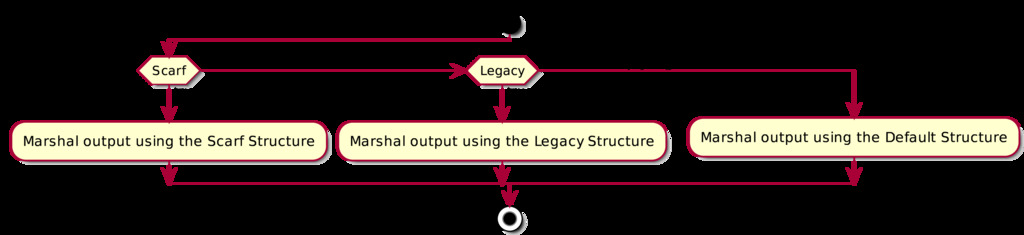}
	\end{center}
	\caption{A very simplistic representation of the CryptoGuard output.}
	\label{fig:crypto_output}
\end{figure}

We redesigned the output structure of CryptoGuard to remain loosely coupled while still being compliant to the SWAMPs Schema.
Following in the popular design patterns, the output system was structured as a Singleton Pattern, under the Creational Design Pattern within the common object-oriented design patterns (\cite{Tutorialspoint_OO},\cite{ObjectOrientedDesign},\cite{BlackWasp2009},\cite{SpringFrameworkGuru}).
Using an abstracted design to reuse the methods the design of the output structure stores the information in a similar fashion to ensure the reuse of the same methods.
Enforcing this ensures the different output designs can all use the same objects.
We created a loosely coupled, modular, and extensible output design that serves the Scarf Output and other outputs as depicted in figure \ref{fig:crypto_output}.

\vspace*{15pt}

\vspace*{15pt}\begin{figure}[H]
	\begin{center}
		\includegraphics[width=\linewidth]{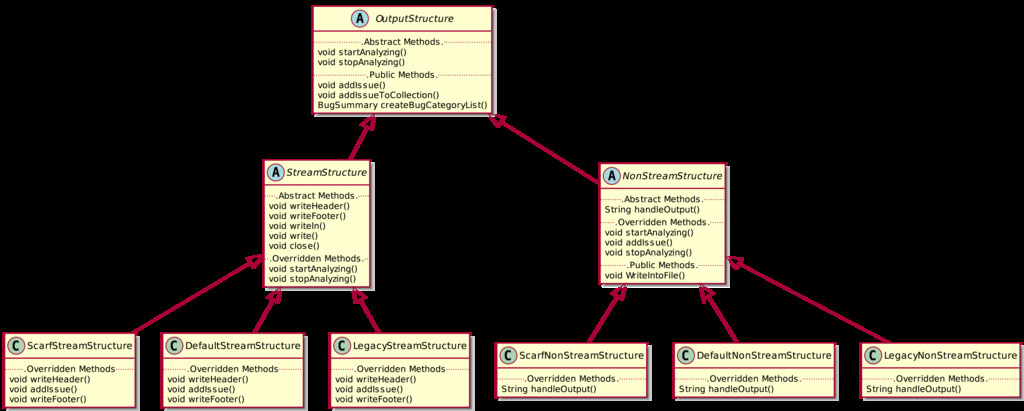}
	\end{center}
	\caption{A simplified representation of the CryptoGuard output abstraction.}
	\label{fig:crypto_output_class}
\end{figure}

In keeping the output system as modular and reusable as possible, we abstracted the marshalling system from the top down as shown in figure \ref{fig:crypto_output_class}; akin to a Factory Design Pattern.
Using this modularization, the rest of the system needs to make use of the 3 methods provided by Structure (startAnalyzing, stopAnalyzing, and addIssue).
This structure efficiently delegates the steps to recreate the output document structure while hiding the business logic.
This structure works for both streaming and non-streaming requirements.

\subsection{Streaming Structure} \label{sse:streaming}

Since SWAMP was potentially scanning monolithic or big projects, they asked that CryptoGuard supported streaming the output.
This also ensures a lower memory used by the JVM during usage, alleviating the memory capacity for larger projects.
Fortunately, before this request, half of the system was already in place (as shown in figure \ref{fig:crypto_output}).
Thus, adding the streaming part of the project only required adding a new top level to the output system and merging the portions under one abstraction.
Encompassing the logic to handle both streaming and non-streaming requests, the three methods (startAnalyzing, stopAnalyzing, and addIssue) readily distribute delegation based on the input arguments.

\vspace*{15pt}\begin{figure}[H]
\begin{lstlisting}[language=java,  firstline=86, lastline=104]
  public void addIssue(AnalysisIssue issue) throws ExceptionHandler {
    super.addIssue(issue);

    log.debug("Marshalling and writing the issue: " + issue.getInfo());
    BugInstance instance =
        marshalling(
            issue,
            super.getCwes(),
            super.getSource().getFileOutName(),
            getId(),
            this.buildId,
            this.xPath);

    String xml = JacksonSerializer.serialize(instance, true, Listing.ScarfXML.getJacksonType());

    if (!xml.endsWith("/>")) this.write(xml);

    // endregion
  }
\end{lstlisting}
	\caption{Streaming the ScarfXML format output.}
	\label{fig:Stream_Scarf}
\end{figure}

To enable this though, only portions of the output structure are rendered in the format in memory (line \# 6), as depicted in the code listing \ref{fig:Stream_Scarf}.
The output structure lightly manipulates each part of the overall format structure to ensure it correctly confirms against its format type.
Removing the empty XML tags ensures the fully rendered file is verified against the ScarfXML Schema.
The structure writes the output only if there is information ensuring there is no empty XML tag (line \# 8).
Empty XML tags are not valid within XML schemas and breaks the validation.

\chapter{Understanding Java Developers Needs} \label{ch:survey}
\section{Understanding the Java Developer Surveys} \label{se:DomainRep}
\vspace*{15pt}\begin{figure}[H]
	\begin{center}
		\includegraphics[width=\textwidth]{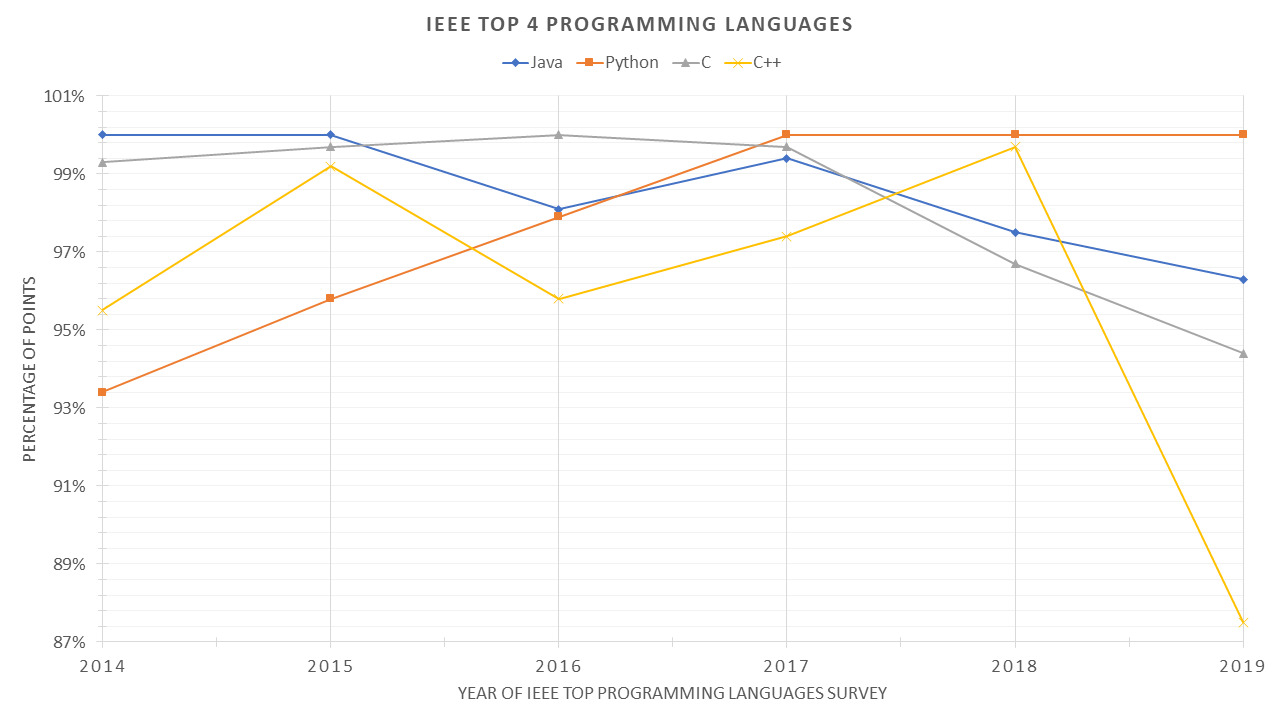}
	\end{center}
	\caption{The IEEE top four programming languages for the past 5 years.}
	\label{fig:IEEETopLang}
\end{figure}

Despite transitioning from a feature-based release schedule to a time-based release schedule with Version 9 in 2017 \cite{Oracle2020}, the community behind Java is still striving.
Though Java has steadily been increasing it is major versions, (as of this writing Java will be on Java Version 14), most users seem to solely use Java 8, the last Long-Term Support (LTS) version from the legacy release schedule.
From Javas first release (as project Oak) in 1995 (\cite{Oraclea}, \cite{Javatpoint2018}), it has historically created a major release every 2 to 3 years \cite{Javatpoint2018}, with the exceptions being Java 7 releasing after 5 years \cite{Javatpoint2018}.
Potentially hindering Javas popularity Oracle, who has been up until recently the main management force creating releases, has let the OpenJDK project take over management \cite{OpenJDK2010} and only supports the LTS versions for 3 years each \cite{Oracle2020_Roadmap}.
Still being a very active and publicly used tool despite it is major release cycle change, the well-known IEEE organization has ranked Java favorably within their top programming language rankings (\cite{Spectrum/IEEE2016}, \cite{Spectrum/IEEE2018}).
Java has been in the top 4 programming languages for the last 5 years as depicted from figure \ref{fig:IEEETopLang}.






Purposed for various technological aspects reaching out to Java developers can be tedious.
This programming language encompasses several domains and various Operating Systems spreading its usage.
Considering many public surveys from predominant companies and Java proponents guides the enhancements to ensure CryptoGuard\cite{Rahaman2019} reaches the largest target audience.
These surveys amalgamated instead of individually to help avoid any bias from within any survey bias.
Over 58,000 people responded over to the surveys over a period of 3 years ensures a wide coverage of Java developers.
The companies creating these surveys are some of the best organizations within the community.
Each of their contributions help to shape the Java community, whether it is by creating significant IDEs, creating guiding documents to better project the community, or hosting fundamental projects, their impact is indomitably.
It is important to note that these survey results were all retrieved from the sources openly published and the results from the participants are self-reported.

We will explain the organizations listed below whose surveys provide great contribution and evidence of the community.
\begin{enumerate}
	\item Snyk
	\item Jakarta EE
	\item JetBrains
\end{enumerate}

\subsection{Information Retrieved from Surveys} \label{sse:SurveyAspects}

Each of the companies providing surveys asked their own set of questions.
The companies questions ask about various aspects of the developers lives.
These questions focus on developer attributes, such as what continuous integration tools are used or their job titles.
Most of their questions focus on Java specific attributes, including the Java version and which Java build tool is used.
These surveys also ask the developers less technical and more personal information.
The less technical questions vary from asking if the developer prefers dogs or cats to if the developer is left or right handed.
I will report primarily on the Java specific information retrieved.
I will also accumulate the commonly asked questions between the surveys during similar survey times.

\section{Snyk} \label{se:Snyk}

\subsection{Background} \label{sse:SynkIntro}

Being a security focused company, Snyk provides a platform for developers to scan both their open-source projects and any containers\footnote{Containers and Kubernetes} they use.
Beyond that, they host an internal database in which they claim to publish NPM vulnerabilities approximately 3 months ahead of the NPM audit.
Collaborating with academia labs, they were able to publish 360 vulnerabilities in 2018.
Not limited to working with academia, they were also able to publish 500 vulnerabilities in 2018 with their own proprietary research.

\subsection{Surveys by Snyk} \label{sse:SynkSurveys}

Ensuring the health of one of the top four languages (shown in figure \ref{fig:IEEETopLang}), Snyk created a developer survey in 2017 and 2019.
Snyks survey coverage describes more information about the developers of the Java language.
We refined and accumulated for the release in the next year respectively.
Combining metrics and demographics between the two surveys, we will only cover the union of the data available between the two.
Working with various Java communities such as (but not limited to) Virtual JUG, Java Magazine, Adopt OpenJDK, and JFokus, they were able to reach out to more than 10,2000 \cite{Vermeer2020} and 2,000 \cite{Maple2018} participants respectively.
Reaching a worldwide demographic, majority (at least 10 percent between surveys) of participants were either from North America or Europe.
More than half of the overall participants (between the 2018 and 2019 surveys) self-reported themselves as Software Developers.

\vspace*{15pt}\begin{figure}[H]
	\begin{center}
		\includegraphics[width=\textwidth]{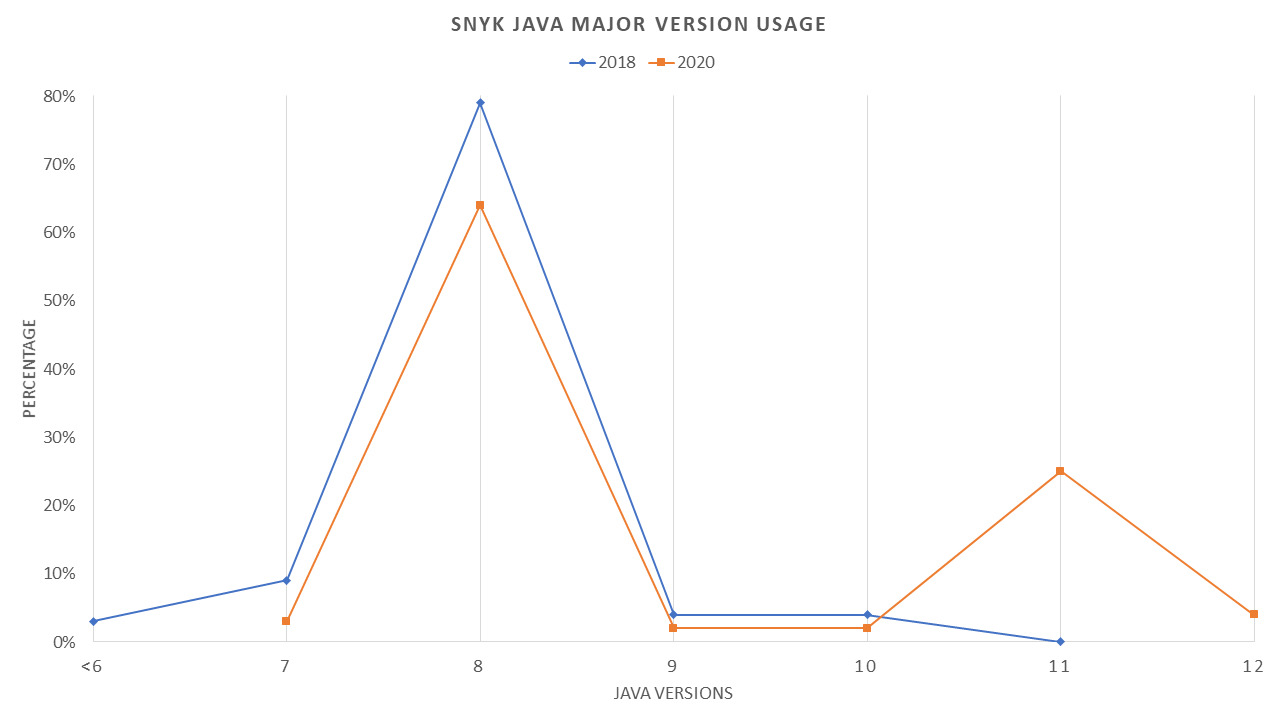}
	\end{center}
	\caption{The Java Version Comparison by Snyk.}
	\label{fig:Snyk_JavaVersion}
\end{figure}

From 2017 to 2019, the use of Java 8 in production being the predominantly used version, dropped from 80 percent to 66 percent \cite{Vermeer2020}, as displayed from figure \ref{fig:Snyk_JavaVersion}.
Java version 7 and 11 seconded version 8 usage\footnote{Notably, Java 7 and 11 are both LTS versions} at the time of each survey respectively.
The decrease in Java 8 usage is likely due to the increase of available Java Versions at the time\footnote{Java 10 was the latest Major Version released for the first survey; Java 12 was the latest Major Version released for the second survey.}.
Though the latest LTS version is 11, majority of developers from this survey are still using Java 8.

\vspace*{15pt}\begin{figure}[H]
	\begin{center}
		\includegraphics[width=\textwidth]{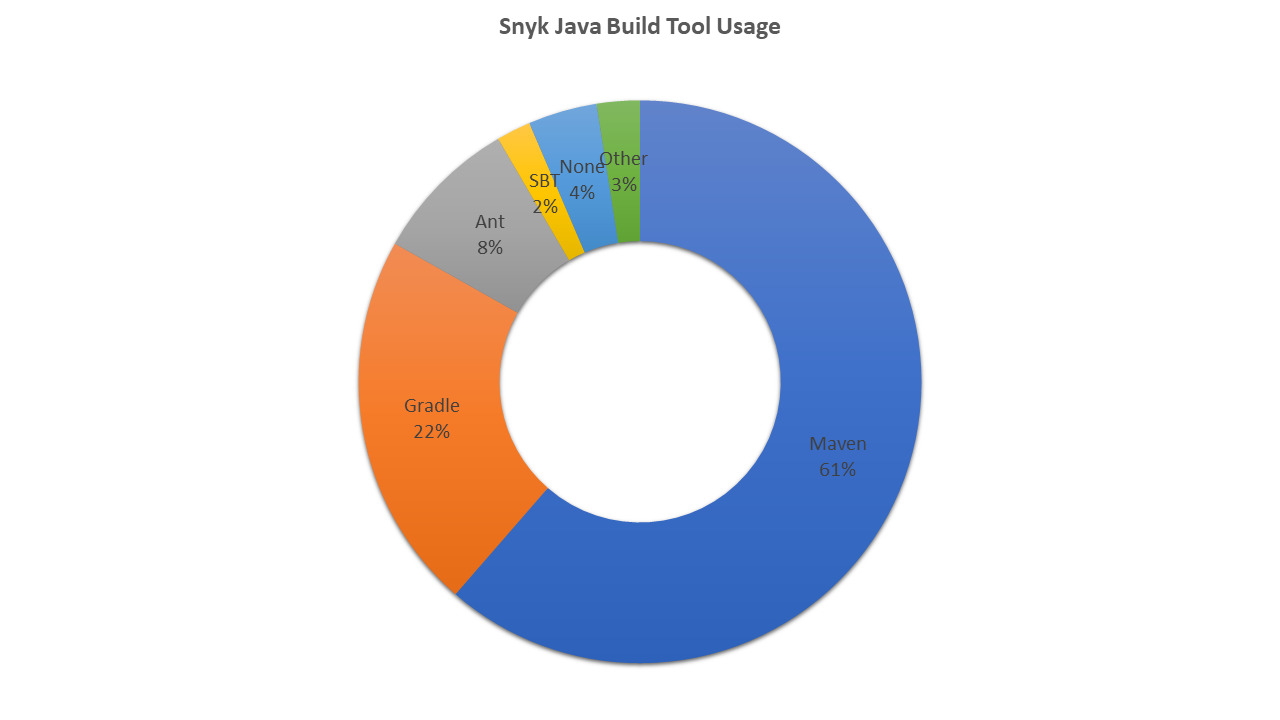}
	\end{center}
	\caption{The Java Build Tool Comparison by Snyk.}
	\label{fig:Snyk_BuildToolUsage}
\end{figure}


Furthermore, within build tools for Java, Maven is the majority build tool of any plugins by at least 39 percent (as represented in figure \ref{fig:Snyk_BuildToolUsage}).
This clear increase is potentially due to the 8 years between Mavens Version 1.0 release \cite{Apache2019} and Gradles Version 1.0 release \cite{Gradle2020}.
Though the Maven covers 61 percent or 7442 participants, we aimed to make the coverage bigger and easily encompass as many developers as possible.
Including the Gradle build tool as well (the second highest build tool according to Snyk), this will increase the range to 83 percent, or 10,126 participants.

\section{Jakarta EE} \label{se:JakartaEE}


\subsection{Background} \label{sse:JakIntro}

Since August of 2017 \cite{EclipseFoundation}, Oracle made a step towards open sourcing Java Enterprise Edition (Java EE) and started creating an open source foundation within the Eclipse Foundation.
Though the project in its entirety (with certain allowances by Oracle) was transitioning to open source, Oracle intended on keeping the name Java EE and the foundation decided on the new name Jakarta Enterprise Edition (Jakarta EE).
Through it is existence, Jakarta EE has helped migrate the Java EE program to ensure it is fully compliant with Oracle and it is fully open source.
Continuing forward, the foundation is handles creating the Java Enterprise Edition (certified via the Technology Compatibility Kit).

\subsection{Surveys by Jakarta EE} \label{sse:JakSurveys}

Within the years of 2018 and 2019, they polled their communities (through social media groups and various user groups) to reach out and better understand their ecosystem.
Throughout both years, the surveys were able to reach out to 3,577 participants.
Like the demographic reached in figure \ref{se:Snyk}, most of the participants were from Europe, Middle East, and Africa \cite{EclipseFoundation2018}, \cite{Milinkovich2019}.
Having a more targeted or more mature audience, these surveys reached most Senior Developers with 41 percent overall\footnote{43 percent in 2018 and 38 percent}.
Using this we can ensure the information we are using also encompasses more mature Java developers.


\vspace*{15pt}\begin{figure}[H]
	\begin{center}
		\includegraphics[width=\textwidth]{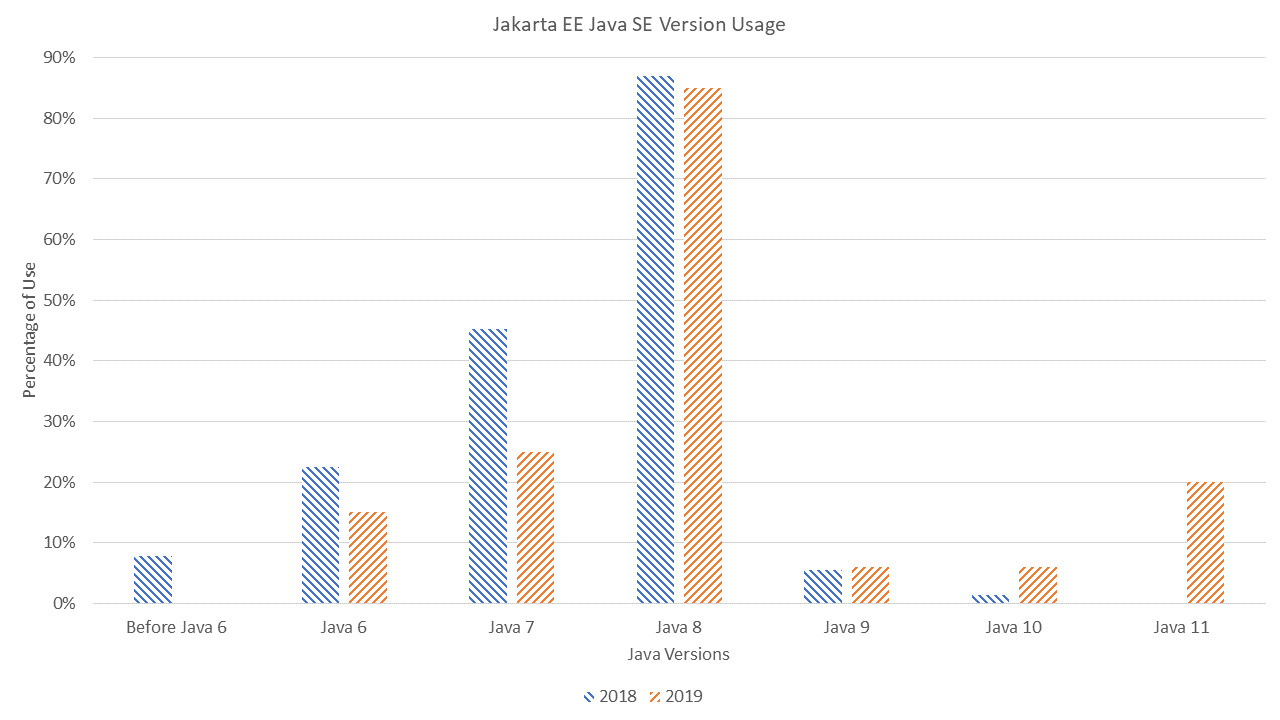}
	\end{center}
	\caption{The Java Version Comparison by Jakarta EE.}
	\label{fig:Jakarta_JavaVersion}
\end{figure}

Retrieving the majority information from both surveys, Java 8 is the majority used version with an overall usage of 86 percent\footnote{86.87 percent in 2018 and 85 percent in 2019}, as shown in figure \ref{fig:Jakarta_JavaVersion}.
Jakarta EE did not measure Java version 12 due to the release date within the duration of this survey.
Despite this caveat since the non-LTS versions (being 9 and 10) both released to less than 10 percent of usage each, Java 12 (being a non-LTS) would obtain the same kind of usage.


\section{JetBrains} \label{se:JetBrains}

\subsection{Background} \label{sse:IntelliJIntro}

As a maker of a few of the most prevalent Integrated Development Environments (IDE), most developers will know of JetBrains, or at least their Java IDE IntelliJ.
With a large acknowledgement in the developer ecosystem more than 8 million developers use JetBrains tools\cite{JetBrains}.
JetBrains has also created their own Java Virtual Machine (JVM) language known as Kotlin in addition to IDEs.
Ranking at 24 in the IEEE 2019 Top Programming Languages \cite{Spectrum/IEEE2019}, the language has earned its solid community reaching a score above other JVM based languages (such as Clojure and Groovy).
Ensuring they know exactly how to improve their IDE and their own language, JetBrains has also created surveys to better encompass the domain of their target audience.
The JetBrains ``State of Developer Ecosystem'' is a survey the company has been running since 2017 to capture various information about the developer landscape.
Fortunately, all their raw data from their surveys was made publicly available {\cite{Chumak2017}, \cite{Chumak2018}, \cite{Chumak2019}},
All of the key insights are taken from combining all the common features (or table labels) and were examined within Microsoft Excel.
JetBrains Surveys release the (anonymized) raw recipient results instead of just the summarization allowing further insight into the results.

\subsection{Surveys by JetBrains} \label{sse:IntelliJSurveys}

\vspace*{15pt}\begin{figure}[H]
	\begin{center}
		\includegraphics[width=\textwidth]{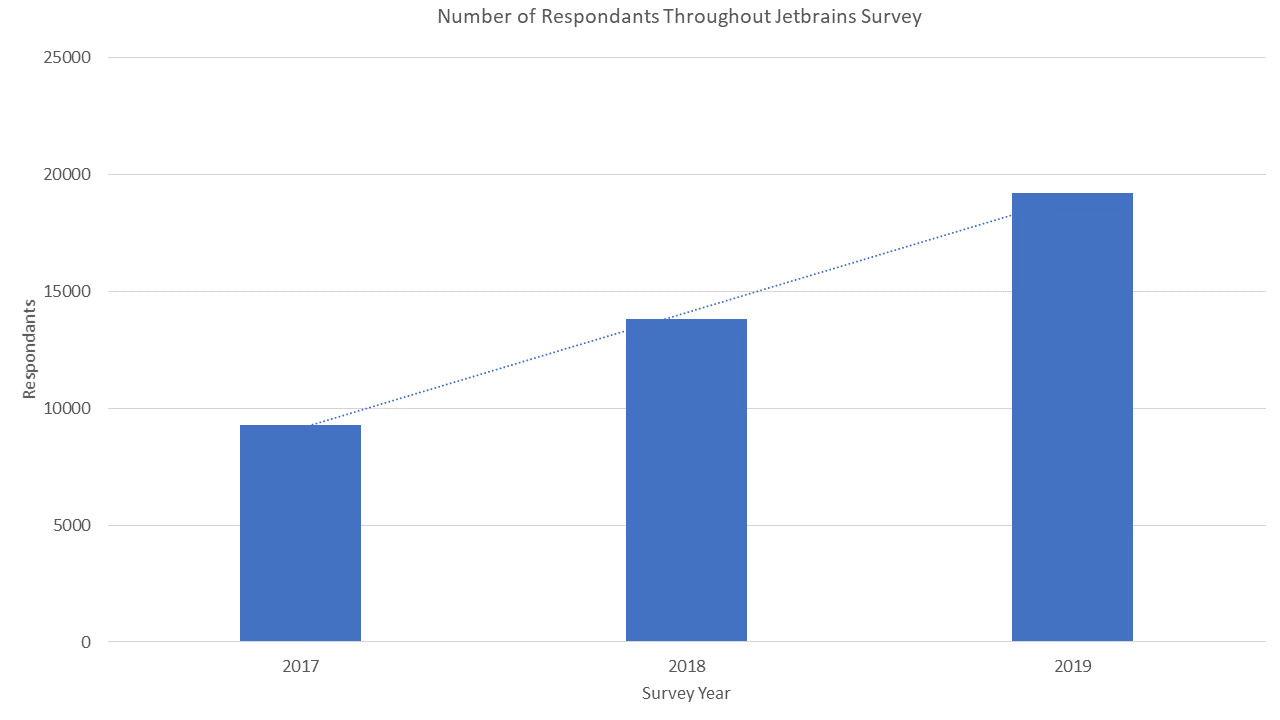}
	\end{center}
	\caption{Number of Respondents from the JetBrains surveys.}
	\label{fig:JetBrains_Users}
\end{figure}

Shown in figure \ref{fig:JetBrains_Users} are the total amounts of participants who the survey each year.
Steadily increasing in participants from the first survey (reaching over 9,200 users) to the final survey (reaching more than 19,000 users), user participation increased 206 percent.
Within this increase, there have been over 42,000 responses within these survey results.
Not only is this survey incredibly expansive in its quantity, the range of participation demographic creates an excellent quality within this data set.

\vspace*{15pt}\begin{figure}[H]
	\begin{center}
		\includegraphics[width=\textwidth]{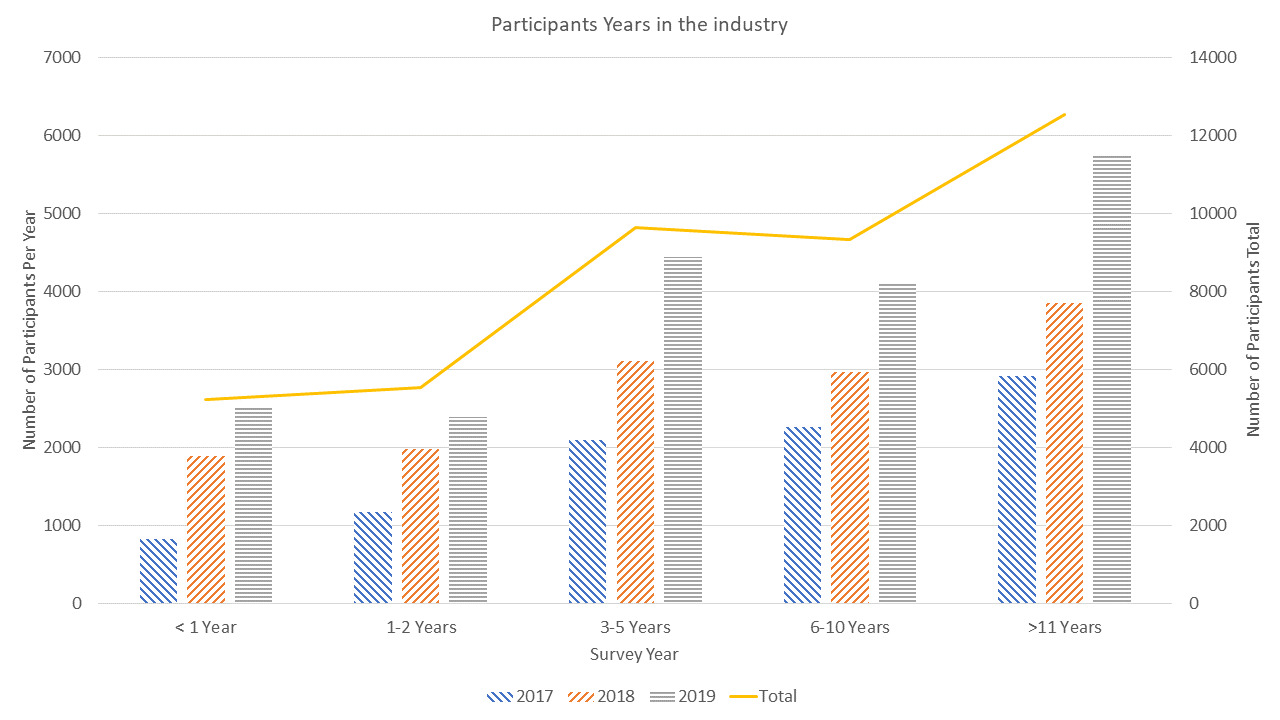}
	\end{center}
	\caption{The years of industry across the participants in the JetBrains survey.}
	\label{fig:JetBrains_UsersExperience}
\end{figure}

Described in figure \ref{fig:JetBrains_UsersExperience}, is the years of experience the participants reported throughout the survey.
With a median of at least 3-5 years in the industry\footnote{3-5 years for year 2018, 2-10 for year 2017 and 2019} this survey represents more experienced developers.
Given most of the developers are more experienced than not, it should be self-explanatory why the average is so high at the range of 6-10 years of experience.
Given this strong weight towards the more experienced developers, these respondents represent the more professional and the more integral developers of companies.

\vspace*{15pt}\begin{figure}[H]
	\begin{center}
		\includegraphics[width=\textwidth]{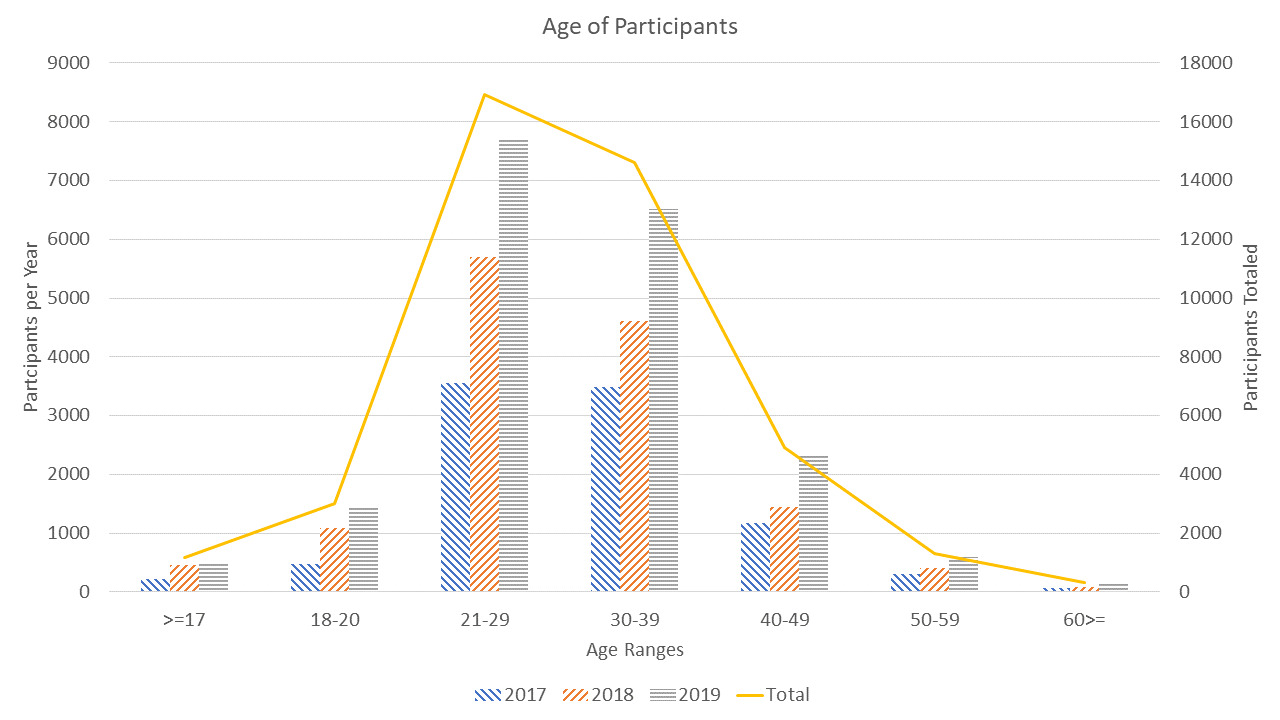}
	\end{center}
	\caption{The age range of the JetBrains survey participants.}
	\label{fig:JetBrains_UsersAge}
\end{figure}

Shown above in figure \ref{fig:JetBrains_UsersAge} JetBrains surveys also represents a vast range of age throughout the developer's participation.
Majority of the age of the participants is between 21-29 and 30-39 group ranges, representing at least 74 percent of the total respondents, with the mean for each year being in the 30-39 year age group range\footnote{75.71 percent for 2017, 74.74 percent for 2018, and 74.07 percent for 2019}.
With a weighted mean at the 30-39 age range, the mode for each year is located within the 21-29 age range.
The median for both 2018 and 2019 both rest at the 21-29 age range, while 2017 respondents rest at the 30-39 age range.

With most of the data centering within the 21-39 age range, it is reasonable to assume most of the respondents are not at the start of their career.
Since the average age range lower bound is typically associated with at least entering the work force after college or being in the work force with a few years' internship, most of the participants are being assumed to be at least young professionals.
Instead of only being able to study young professionals or students with a various range of experience, with this study we are analyzing data from relatively more mature professionals who are used to working at a company with live projects.
This group of participants also creates a good opportunity to measure companies' standards, since it is likely young professionals are being trained and well versed with the programming and coding standards of their team, and subsequently their company.

\vspace*{15pt}\begin{figure}[H]
	\begin{center}
		\includegraphics[width=\textwidth]{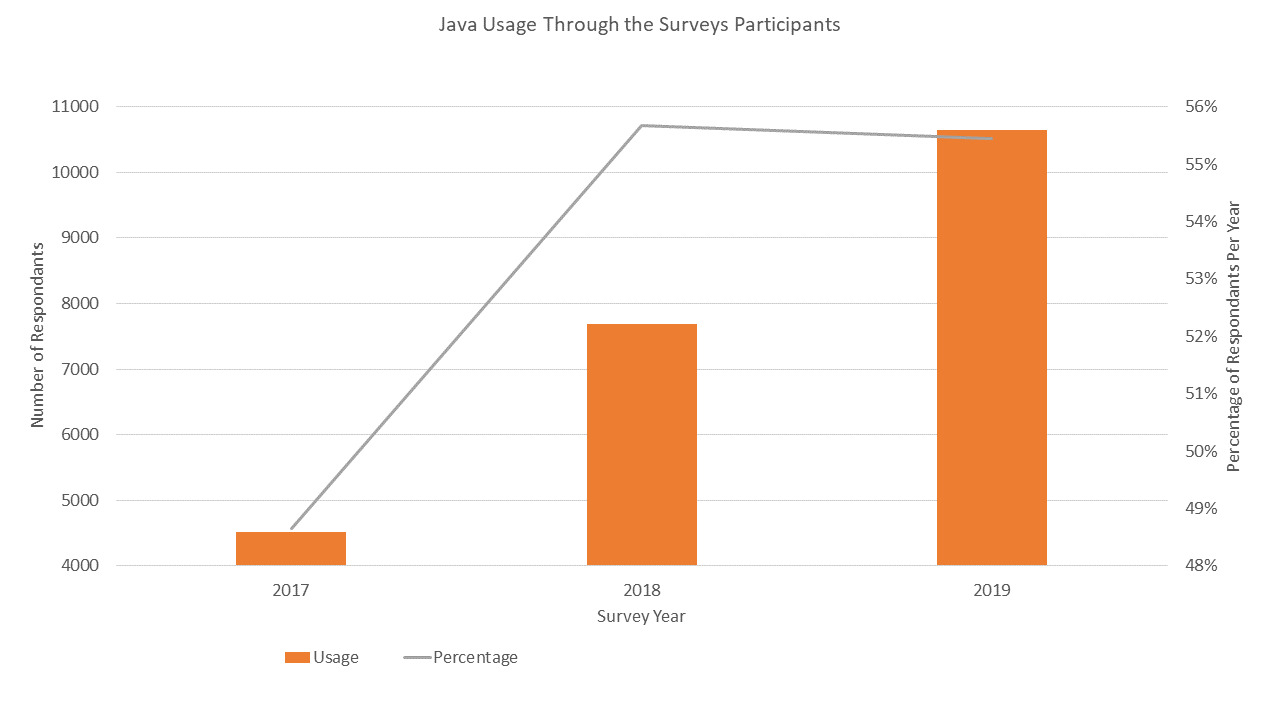}
	\end{center}
	\caption{The number of JetBrains survey respondents who use Java.}
	\label{fig:JetBrains_JavaUsers}
\end{figure}

Since JetBrains supplies a plethora of IDEs based around various programming languages, the participants are not only Java developers, but developers from using a vast number of tools.
Depicted in figure \ref{fig:JetBrains_JavaUsers}, is the number of developers who use Java as either their primary or secondary language.
Though there is a steady increase of Java developers per each survey (represented by each bar corresponding to the year), the percentage line represents how the relative amount of users per year (the grey percent line with a separate y-axis bar on the right).
In short, the percentage line measures the Java developers in relative to the number of people in that specific survey year; for example, the 2017 survey had 4,520 respondents who use Java, however taken out of the total respondents that year (9292) represents approximately 48 percent of the participants for that year.
Despite there being a slight percentage decrease of Java developers from 2018 to 2019\footnote{a 22 percent drop}, the overall weighted average of Java developers is 54 percent.
Though not an entirely targeted survey, any Java specific metrics taken throughout the survey will explicitly include Java consumers (using Java as a primary language or a secondary language) to ensure the information we retrieve is directly usable for our purposes.

\begin{figure}[H]
	\begin{center}
		\includegraphics[width=\textwidth]{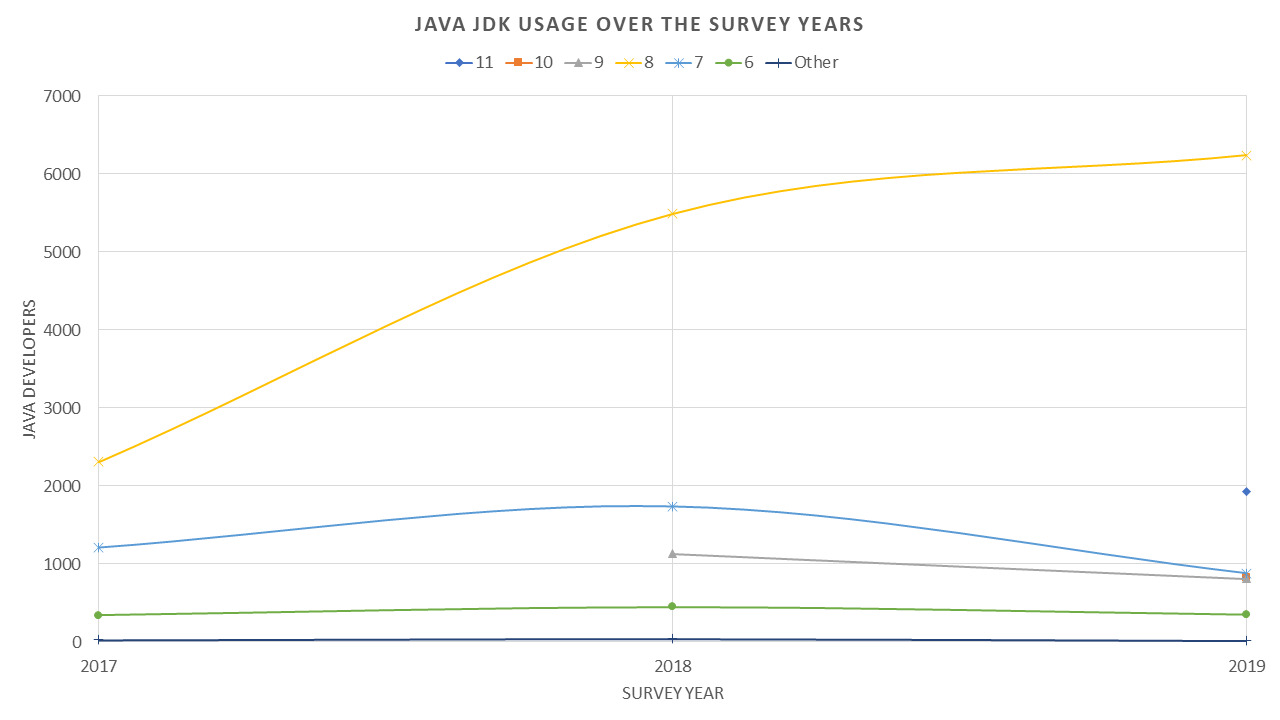}
	\end{center}
	\caption{Java JDK comparison by JetBrains respondents.}
	\label{fig:JetBrains_JavaVersion}
\end{figure}

\begin{figure}[H]
	\begin{center}
		\includegraphics[width=\textwidth]{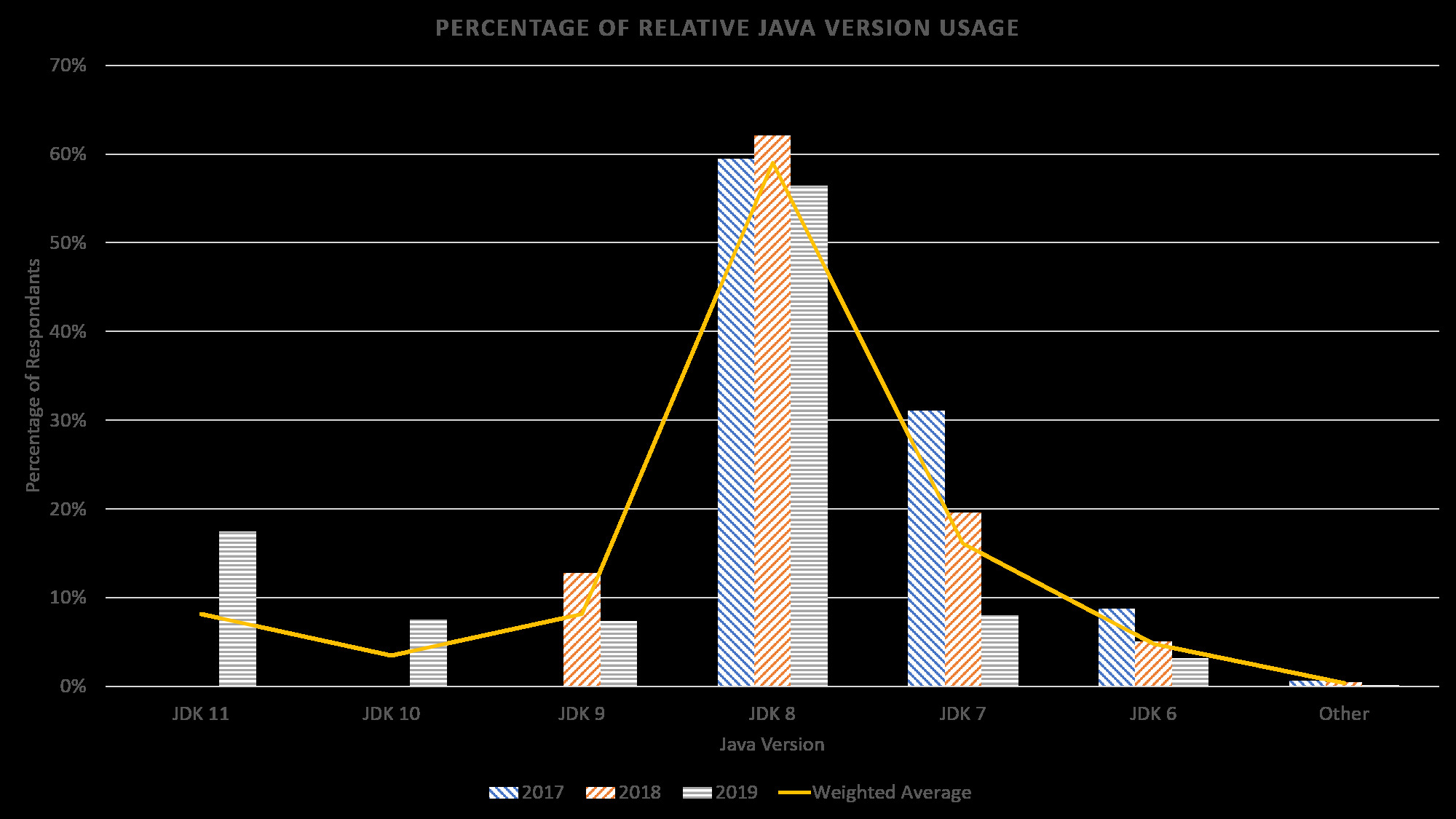}
	\end{center}
	\caption{Java JDK relative comparison from JetBrains respondents.}
	\label{fig:JetBrains_RelativeJavaVersion}
\end{figure}

Similar to earlier surveys (\ref{sse:SynkSurveys}, \ref{sse:JakSurveys}), majority of the Java consumers are still actively using Java 8 as shown in figure \ref{fig:JetBrains_JavaVersion}.
Though Java versions 11, 10 and 9 usage\footnote{except for Java 9 in 2018} are not directly comparable throughout the surveys due to their release dates releasing during or after the survey, there is a clear lead of Java 8.
Throughout all the surveys, the closest Java version competitor is Java 7 in 2017, with only a 28.34 percent difference.
Throughout the other surveys, the difference increases from 2018 to 2019 in 48.70 percent to 57.56 percent.
Noticeably, even when Java 11 (the next LTS) Release came out the adoption right at the time of taking the survey was only 25.79 percent.
OpenJDK released Java 11 in September 25, 2018\cite{OpenJDK} (for general availability) so developers had at least 3 months before the survey opened.
Further exemplified in figure \ref{fig:JetBrains_RelativeJavaVersion} is the usage discrepancy of the overall usage of each Java Version.
We created the weighted average by determining first the number of respondents who answered this survey question, then finding and combining the weighted percentage relative to each of their respective years, then finally creating the graph representation.
Of all the 19,086 respondents who responded to questions determining which Java Version they use, 73.57 percent or 14,041 participants use Java 8 in some capacity.
Given that the survey is representative of more mature developers (displayed from figure \ref{fig:JetBrains_UsersExperience}) it is certain that they use Java 8 the most.

\begin{figure}[H]
	\begin{center}
		\includegraphics[width=\textwidth]{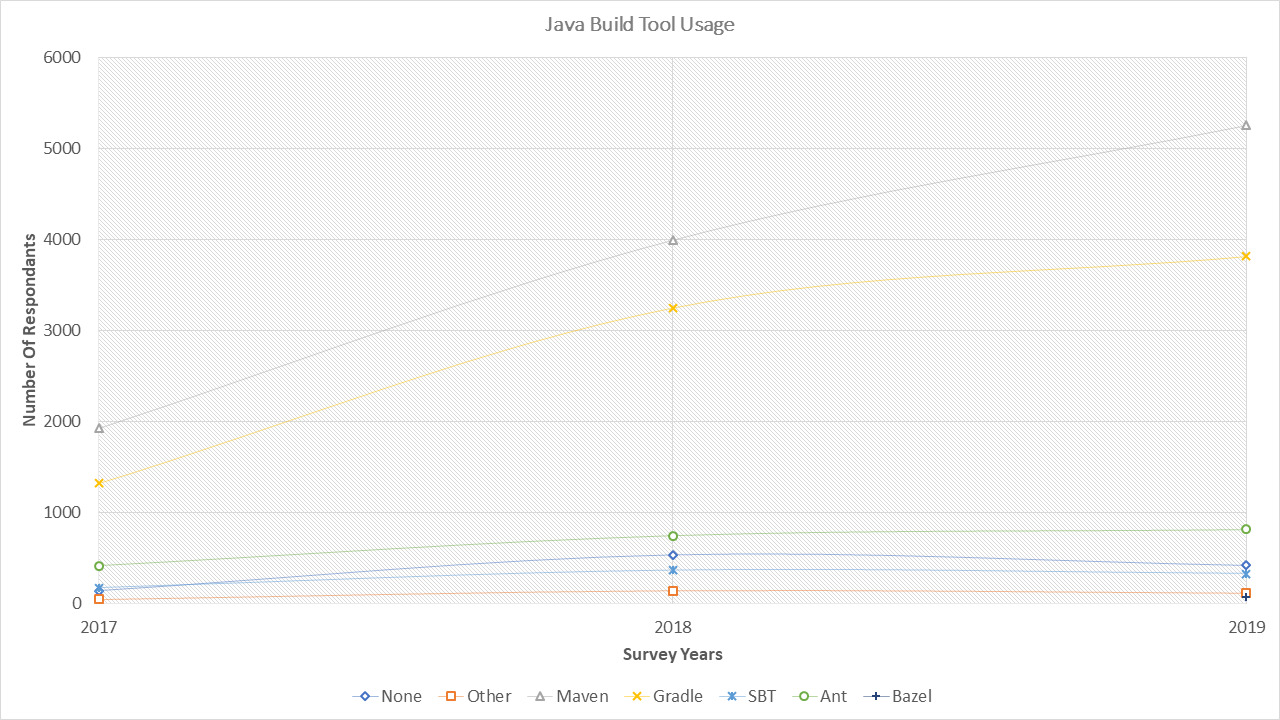}
	\end{center}
	\caption{Java build tool comparison from JetBrains respondents.}
	\label{fig:JetBrains_BuildTool}
\end{figure}

\begin{figure}[H]
	\begin{center}
		\includegraphics[width=\textwidth]{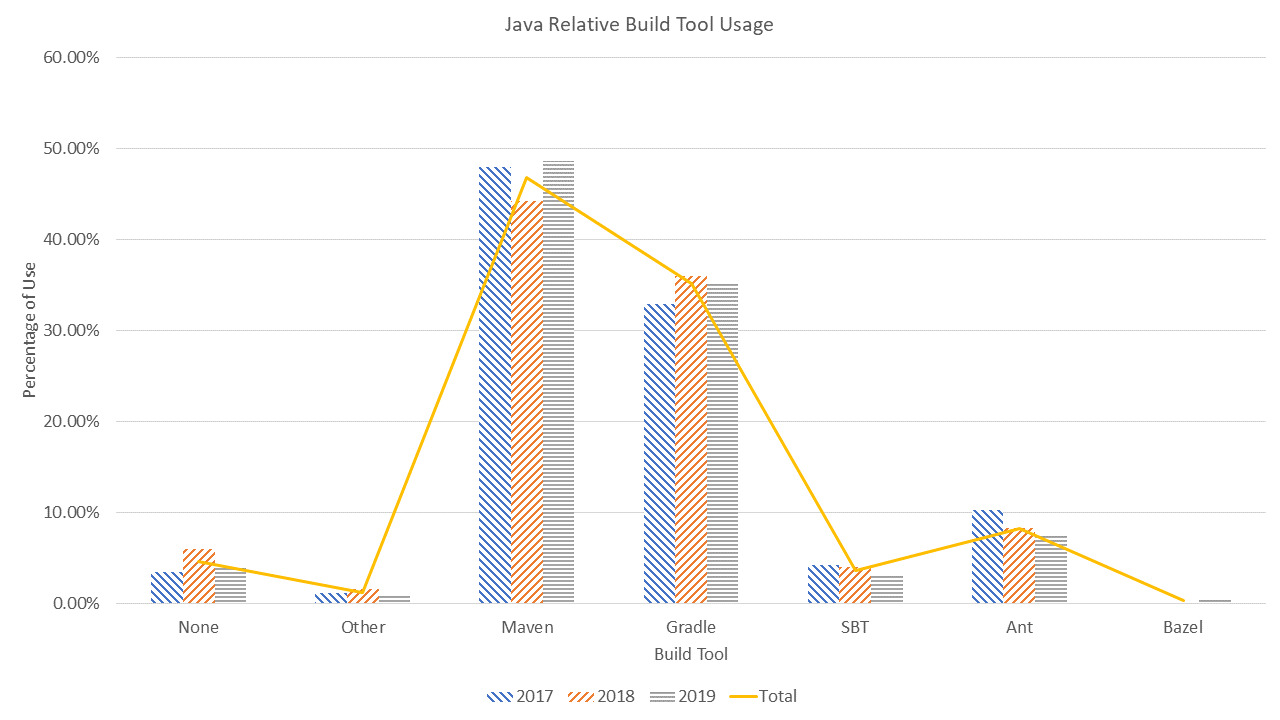}
	\end{center}
	\caption{Java build tool relative comparison from JetBrains respondents.}
	\label{fig:JetBrains_RelativeBuildTool}
\end{figure}

Since normal compilation throughout projects can become quite complicated and exhaustive, there has appeared certain build tools that are known throughout the developer ecosystems.
Build tools have been created primarily at aiding the development process in numerous ways, usually all which helping to aid the compilation process.
Build tools can also help with (but are not limited to) automatically running tests on source code, deploying artifacts or packages to specified locations, automatically retrieving dependencies, or creating a release candidate pertaining for a certain specification.
One of the overall benefits build tools provides is a standard that projects can follow and adopt to automate whatever is available via the build tool.
Universally recognized examples of build tools include Makefile and CMake for C or C++ projects, pip and pip3 for Python projects, Cargo for Rust projects, and NPM for JavaScript projects.

Not only allowing us to study the statistics over the span of three years and combine certain metrics, represented in figure \ref{fig:JetBrains_BuildTool} are the popular Java build tools listed from the respondents.
The top two build tools for Java, Maven and Gradle, take up more than 80 percent of the usage.
Rising from the first majority percentage of 80.85 percent to 83.91 percent from 2017 to 2019, it is clear their relevance is only increasing.
Still increasing the percent of usage, about 85 percent of Java Developers (overall) use at least Maven or Gradle as one of their build tools.
With slight decrease in usage from 2018 to 2019, the trend line shows a steady increase from 2017 to 2019 as further depicted in figure \ref{fig:JetBrains_RelativeBuildTool}.
Able to select more than one choice for their build tool of use, from 2017 to 2019 the gap between the top two build tools Maven and Gradle increased.
Though with a close difference between the two, Maven is the most used build tool with 45 percent, 22 percent, and 37 percent increases for each of the 2017, 2018, and 2019 surveys respectively.
Combining these together dictates Maven achieves an overall weighted average increase of 33 percent over Gradle.
This is still likely due to the 8-year date release between Maven and Gradle (\cite{Apache2019}, \cite{Gradle2020}).

\vspace*{15pt}\begin{figure}[H]
	\begin{center}
		\includegraphics[width=\textwidth]{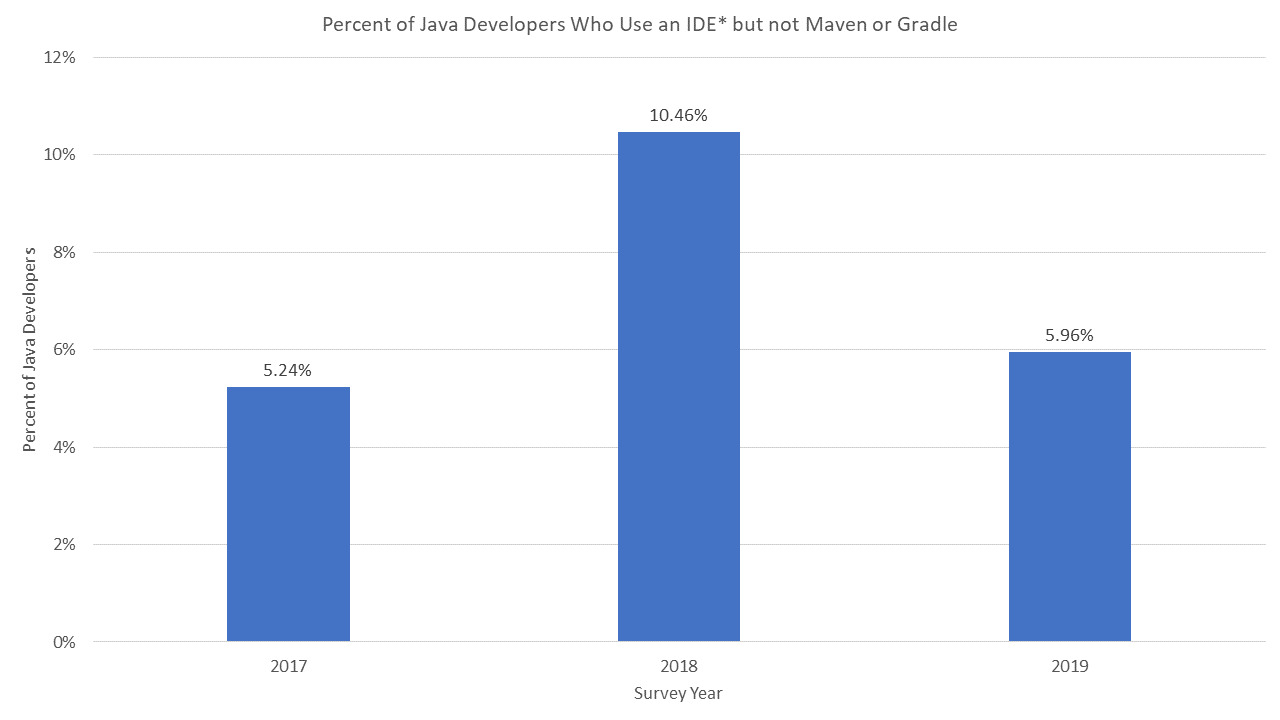}
	\end{center}
	\caption{Percentage of Java Developers who use an IDE but not Maven or Gradle.}
	\label{fig:JetBrains_IDENoMavenGradle}
\end{figure}

JetBrains created the popular Java IDE IntelliJ potentially introducing bias towards their products.
However, with our analysis focusing on the build tool usage and there is plugin support (either official or unofficial) for most of the build tools discussed, our retrieved metrics are unaffected by such bias.
Only due to the availability of the raw respondents' survey data, we can prove with these surveys the build tool coverage despite the numerous types of IDEs.
Expanding on the coverage the usage of either Maven or Gradle provides as a build tool despite the IDE used, shown in figure \ref{fig:JetBrains_IDENoMavenGradle} is the amount of Java Developers who \textbf{use an IDE but not Maven or Gradle}.
Encompassing a wide range of IDEs, included in this category include programs such as JetBrains products, Eclipse based IDEs, NetBeans, Vim, and others.
Able to specifically target the top two build tools throughout the survey, we were able to find that an overall weighted average of 7.33 percent of users who use an IDE do not use Maven or Gradle.
This proves targeting both top two build tools provide a majority coverage to developers despite the specific IDE they use.

\newpage
\vspace*{15pt}\begin{figure}[H]
	\begin{center}
		\includegraphics[width=\textwidth]{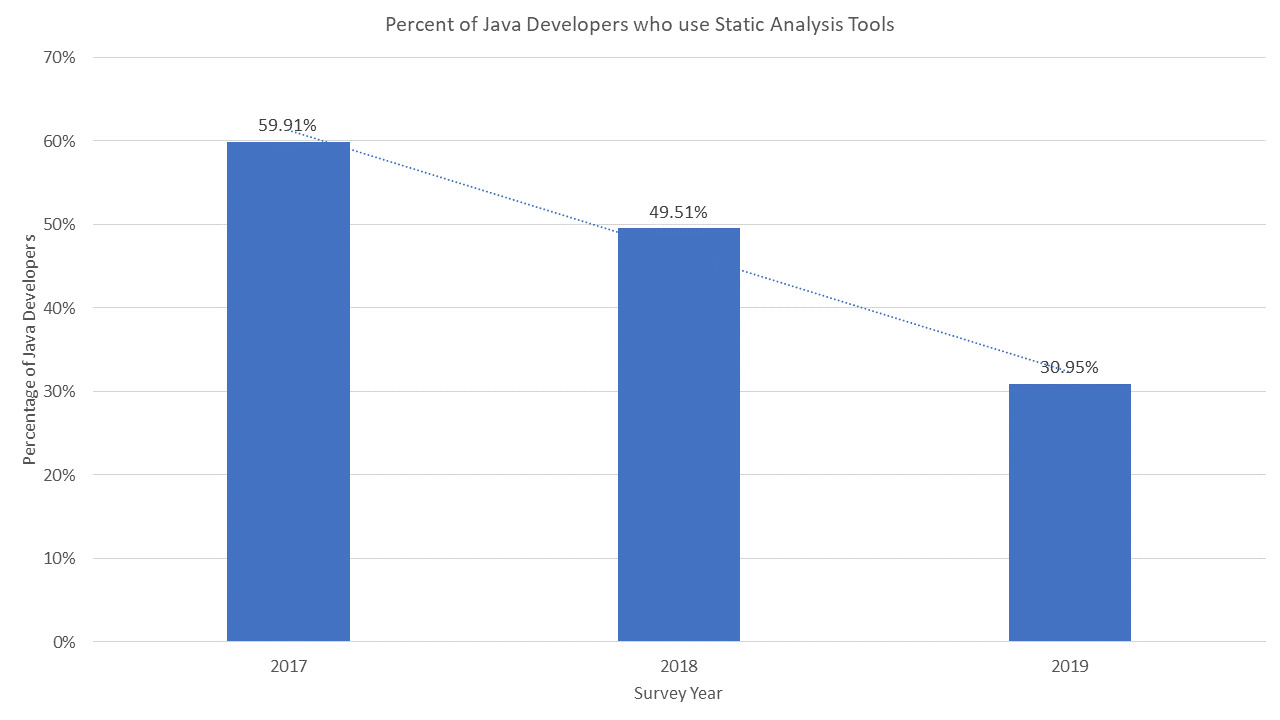}
	\end{center}
	\caption{Percentage of Static Analysis Tools Amongst Java Developers}
	\label{fig:JetBrains_StaticAnalyzer}
\end{figure}

Throughout this survey there was a very indicative question pertaining to the usage of Static Analysis Tools amongst developers.
While the original infographics represent the overall usage amongst the developers surveyed, we were able to explicitly target Java developers and determine they are percent usage of Static Analysis Tools, as displayed in figure \ref{fig:JetBrains_StaticAnalyzer}.
Unfortunately, there is a steady decline in the trend line of usage in 2017 to 2019 despite the increase of participants and spread of demographic.
Similarly, this may be due to the abundance of tools, or the lack of integration with the build system as one of the respondents from \cite{10.5555/2486788.2486877} claimed.
Further among that point, 19 out of 20 of the developers surveyed exclaimed the importance of workflow integration \cite{10.5555/2486788.2486877}.
Since build tools like Maven and Gradle can be fully automated in Continuous Integration (CI) pipelines let alone the core process of compilation provides assured integration in opposition of an IDE specific plugin.
Another reason for decrease of usage could be the dependency of online tools such as Stack Overflow.
78 percent of Software Developers \cite{Meraki2019} use Stack Overflow for assistance and community driven assurance that the highest ranked answer is the correct one, given the direct increase of Cryptographic questions posted as described by \cite{meng2018secure}.

\section{Main Findings} \label{se:SurveyGeneral}

Creating an accumulation throughout the surveys taking during the same years reveals similar patterns from the previous individual surveys.
I combined the three sets of surveys from 2018 and 2019, increasing the predominance of patterns.

\begin{figure}[H]
	\begin{center}
		\includegraphics[width=\textwidth]{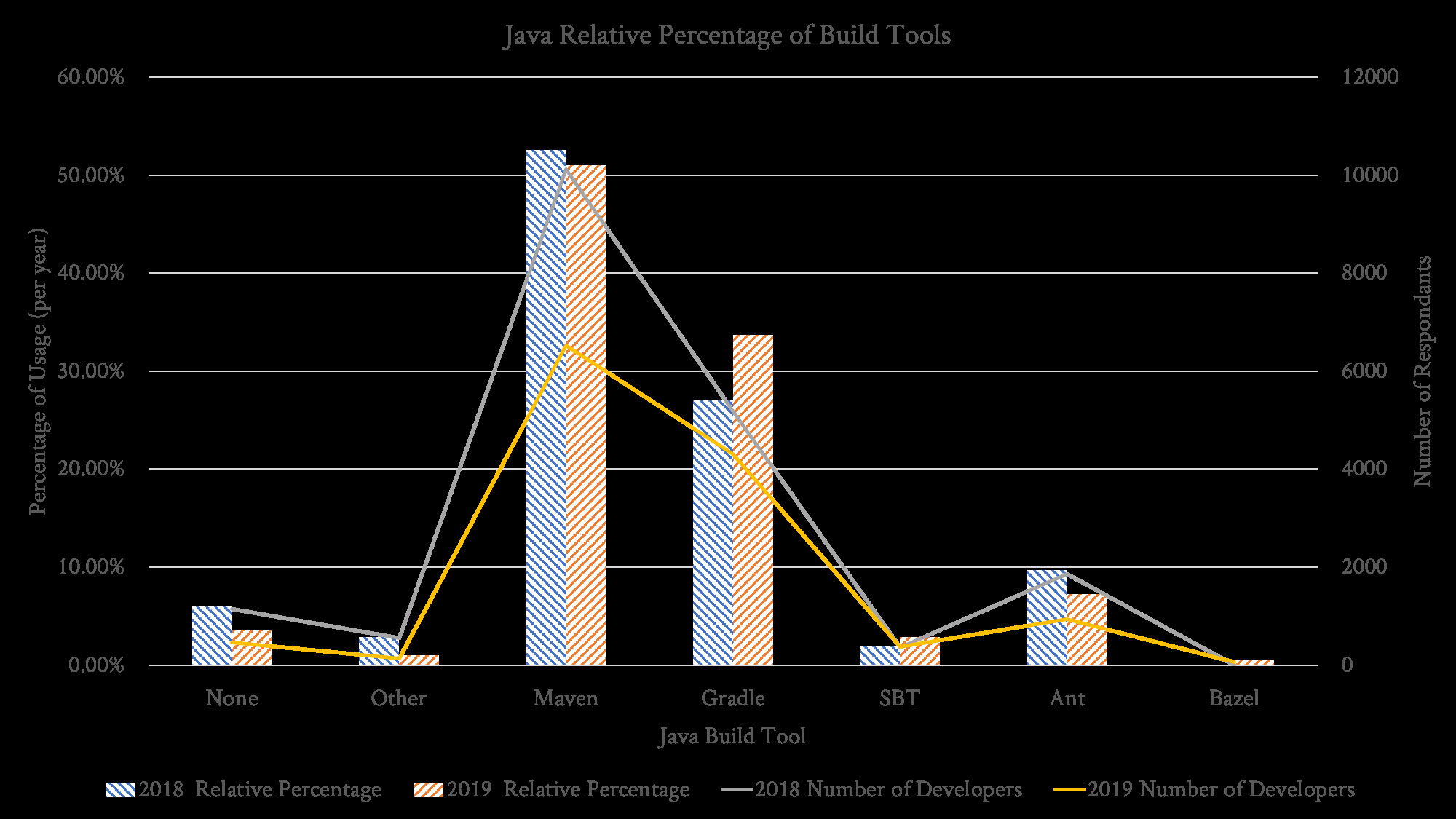}
	\end{center}
	\caption{An accumulation of the Java build tools across the multiple surveys.}
	\label{fig:JetBrains_BuildTools_Tot}
\end{figure}

\begin{figure}[H]
	\begin{center}
		\includegraphics[width=\textwidth]{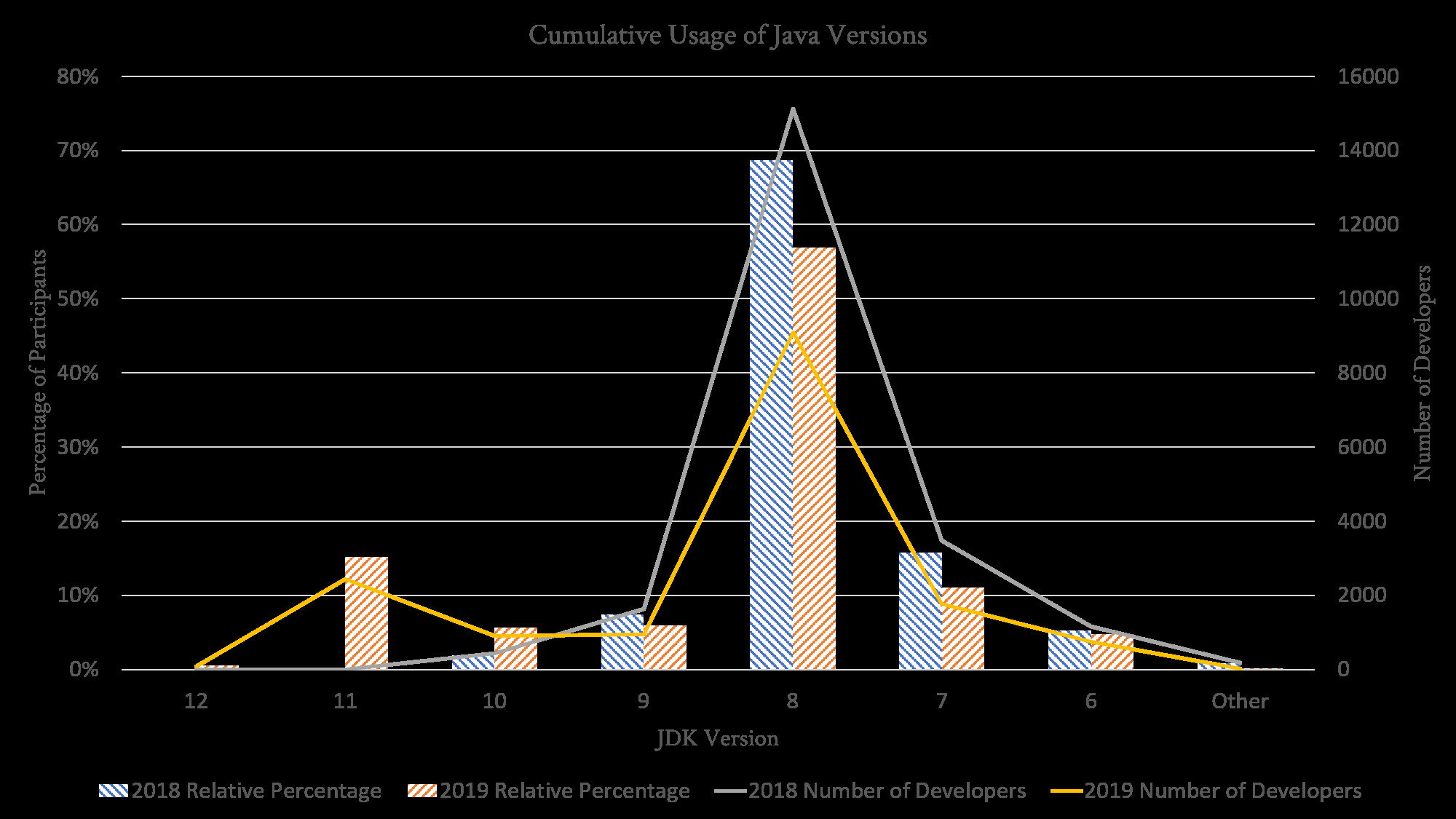}
	\end{center}
	\caption{An accumulation of the Java version usage across the multiple surveys.}
	\label{fig:JetBrains_JavaVersion_Tot}
\end{figure}

Maven and Gradle are the top two build tools used by Java developers, as shown in figure \ref{fig:JetBrains_BuildTools_Tot}.
Using a weighted average per year, there is at least an 11.3 percent difference between Gradle and the next highest used tool.
These results clearly define how Maven and Gradle are the most used build tools by Java developers.
Java developers use Java 8 by a strong majority.
As shown in figure \ref{fig:JetBrains_JavaVersion_Tot}, Java 8 is the highest used version.
Better depicted by the relative percentage, there is an increase of 49.97 percent across all the years.
These results summarize at least 37,000 developers.
With the latest two highest used Java Versions being 11 and 8, developers tend to only used long term support versions.
This is likely due to the non-long term support versions are only supported for 6 months.
After Java 11, the next long term support version is Java 17.
Until Java 17 is released, it is likely the majority of developers use either Java 8 or Java 11.
Since Java 8 is the most used version of Java, CryptoGuard is on the majority used version despite the version.

Taking an overview of all the surveys from Snyk, Jakarta EE, and JetBrains, we were able to determine the following concepts:

\begin{enumerate}
	\item These surveys encompass most Software Developers.
	\item The participants in the survey have a wide demographic, in both country location, age, and years of industry experience.
	\item Java 8 is the most used Java Version JDK \footnote{surpassed by 4 versions at the approximate time of surveys taken} by 49\%\footnote{across 2018 and 2019 of 37,971 developers}.
	\item Maven and Gradle are the top two used build tools by 72\%\footnote{across 2018 and 2019 of 32,030 developers from JetBrains and Snyk}.
\end{enumerate}

From the accessibility of the JetBrains survey and responses, we were able to further delve into the responses and rationalize the following points:

\begin{enumerate}
	\item Static Code Analysis amongst Java Developers has been steadily decreasing by an average of 12\%\footnote{from 2017 to 2019 within JetBrains Survey of 22,841 Java Developers}.
	\item Despite the usage of an IDE, 85\%\footnote{from 2017 to 2019 within JetBrains Survey of 22,841 Java Developers} of Java Developers are still likely to use either Maven or Gradle.
\end{enumerate}

\chapter{Usability Enhancements} \label{ch:enhancements}
\section{Plugins} \label{se:Plugins}

We created multiple plugins to integrate CryptoGuard with the build process of a project.
This creates a one-time setup that allows CryptoGuard to automatically scan the project during compilation.
We created a plugin specifically for both build tools Maven and Gradle.
Utilizing both systems allows a majority coverage of Java Developers, as depicted from figure \ref{fig:JetBrains_RelativeBuildTool} with the sample size of more than 22,000 developers.
In the following section we will describe the efforts taken to integrate CryptoGuard within the targeted Java build tools.

Though there are still people and groups that use each kind of tool, majority of people have been using Maven and to a lesser extent Gradle.
Though there is no perfect statistic to help measure this, the \underline{State of Developer Ecosystem} by JetBrains helps to display a relatively complete depiction of various programming community.
Though the developers may be developers JetBrains products (free or professional based tools) and are biased, there is enough representation to help highlight the popularity of the build tools.
Listed below are the results of the different results for build tools of the last three years.

Showing the trend lines through the time of the three years from the figure \ref{fig:JetBrains_RelativeBuildTool}, the lowest Maven has been within the Survey was 68 percent in 2018.
Though Gradle is the second highest build tool, within the same year (2018) Gradle is at it is highest yet is still 21 percent lower than Maven.
Given the steady trends for the both Maven and Gradle is enough to back the decision to target as the top two build tools utilized for the future.
We created these two plugins to extend CryptoGuard natively into each of the tools to ensure better ease of use.

\vspace*{15pt}\begin{figure}[H]
	\begin{center}
		\includegraphics[width=\linewidth]{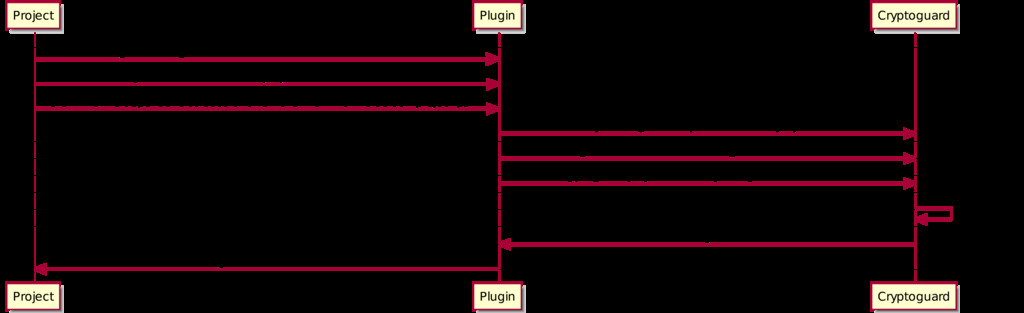}
	\end{center}
	\caption{The general design of CryptoGuard plugins.}
	\label{fig:Plugin}
\end{figure}

Given the nature of library, the plugins only needed to be of a ``wrapper'' type design.
When we extended CryptoGuard to provide an easy entrance, each plugin only needs to handle and transition the proper settings and variables.
The design shown in figure \ref{fig:Plugin}, describing the general message passing between the libraries.

\subsection{Maven Plugin} \label{sse:Plugins_Maven}
Since Maven plugins directly use Java files to interpret the XML based configuration, this plugin type was easier to design and create.
Since the plugin only acts as wrapper for the CryptoGuard library, the plugin itself mostly provides more enhanced configuration abilities.

After handling the XML configuration arguments, this plugin formats them to directly pass to CryptoGuard.
Given the standard maven format of the project, the compilation path and directories are known by default.
Since Maven also acts as a standardized dependency manager, the location of the cached dependencies is known as well.
Through some exploration, it was determined the default location where Maven stores the local dependencies for regardless of operating system.
CryptoGuard creates the path dynamically upon the scan running to ensure it is environment independent.

\subsection{Gradle Plugin} \label{sse:pluginGradle}
Since Gradle uses the dynamic programming language Groovy for compilation, the plugin itself uses both Java and Groovy programming.
Like the Maven Plugin, the plugin itself mostly provides more enhanced configuration abilities to directly pass to CryptoGuard.
The groovy portion is only for handling the various features provided for the configuration script while the core functionality handles the CryptoGuard wrapping.

After extending the CryptoGuard functionality to scan Java Class files, this provided the availability for scanning (since Java Files are still currently unstable).
Given the Gradle format of the project, it is automatically known in compilation path and automatically known to the library.
Since Gradle acts as a dependency manager, the location of the dependency locations is inherently necessary.
Through some exploration, it was determined the default location where Gradle stores the local dependencies.
CryptoGuard dynamically creates the path during the scan running to ensure it is path independent.

\section{CryptoSoule} \label{se:script}
One of the later additions onto the project is a python script that allows a more granular level of help.
Instead of solely extending the standard documentation (via USAGE.md or README.md), we chose a script since it can add more dynamic and detailed information.
We originally created the script to run the integration tests individually (for an isolated environment), the functionality grew from the requests of SWAMP.
It is entirely self-contained and can run either from a clone of the project repository or remotely as a standalone script.

\vspace*{15pt}\begin{figure}[H]
\begin{lstlisting}[language=make, linewidth=\textwidth, firstline=4, lastline=27]
cryptoguard: V04.00.03
Gradle Version: 4.10.3
Java Build Version: 1.8.232
===============================================================================
	./cryptosouple.py rawArgs: Prints the raw arguments of the program.
	./cryptosouple.py exampleArgs: Sample examples of running the program with arguments and explanations.
	./cryptosouple.py writeUsage: Write the example args to a markdown file (USAGE.md).
	./cryptosouple.py checkEnv: Checks (suggestions to set them if missing) the environment variables.
	./cryptosouple.py projectType: Displays some information about the project types available to scan.
	./cryptosouple.py outputType: Displays some information about the various output types available to write out as.
	./cryptosouple.py exceptionType: Displays information about the standardized exceptions.
	./cryptosouple.py clean: Cleans the project.
	./cryptosouple.py build: Builds the project.
	./cryptosouple.py refresh: A shortcut to clean and build the project.
	./cryptosouple.py hash: Determines the hash of a freshly built project.
	./cryptosouple.py buildCmd: Build the command to run, NOTE experimental!
	./cryptosouple.py test: Runs a specified test, can also supply the fullyqualified test names, (i.e. smapleTestOne or sampleTestOne,testCWEListing).
	./cryptosouple.py tests: Runs all of the tests crawled.
	./cryptosouple.py testType: Runs a specified set of tests.
	./cryptosouple.py testsHelp: Shows helpful information about the tests crawled.
	./cryptosouple.py genTest: Generate a test method from a string command. ex: ./cryptosouple.py genTest "-c CMD..." [file]
	./cryptosouple.py displayTests: Displays Tests available.
	./cryptosouple.py help: Displays helpful information to the user if it's the first time running.
	./cryptosouple.py offline: Write enough of the information internally for this script to run stand-alone.
\end{lstlisting}
	\caption{The introduction to running the CryptoSoule program.}
	\label{fig:./CryptoSoule.py Intro}
\end{figure}

Requiring no dependencies and no additional setup necessary, a developer can directly execute this file for help.
Shown within figure \ref{fig:./CryptoSoule.py Intro} are all the available options to the developers.
Due to various requests, most of the requests handle meta information such as creating an executable command line argument, displaying a survey link for the project, and hashing the projects executable.
Unfortunately, due to the complexity within CryptoGuard's main library Soot, multiple environment variables must be explicitly set and will not run if unset.
To help mitigate this and as a direct response to some of SWAMPs requests, the script also provides a command to explicitly state the variables needed and help the user set them.
Though only three environment variables, the script allows the developers to automatically check it and allow the user to set it for their shell session.

\subsection{Interactive Helper} \label{sse:buildScript}
After the beginning of contributions to CryptoGuard, the command line arguments have been increasingly more complicated and growing due to the various requests from SWAMP.
Given the increasing range and various command line traversal paths, we created a Python file to generate an interactive command builder.
Though we originally created the script for running the unit tests in a custom manner, it grew to speed up the developers understanding the command line arguments.

\vspace*{15pt}\begin{figure}[H]
\begin{lstlisting}[language=Java,  linerange=20-40]
  FORMAT(
      "in",
      "format",
      "Required: The format of input you want to scan, available styles "
          + EngineType.retrieveEngineTypeValues()
          + ".",
      "format",
      null,
      true),
  SOURCE(
      "s",
      "file/files/*.in/dir/ClassPathString",
      "Required: The source to be scanned use the absolute path or send all of the source files via the file input.in; ex. find -type f *.java >> input.in.",
      "file(s)/*.in/dir",
      null,
      true),
  DEPENDENCY("d", "dir", "The dependency to be scanned use the relative path.", "dir", null, false),
  OUT(
      "o",
      "file",
      "The file to be created with the output default will be the project name.",
\end{lstlisting}
	\caption{An excerpt of the raw arguments within CryptoGuard.}
	\label{fig:Args Enum}
\end{figure}

Though most of them are flags enabled to allow the consumer to have an easier time with either the runtime or the results, several of the arguments have chained dependencies between each other.
To ensure further clarity for these dependencies and to add a better representation of the options for the user, the interactive command builder runs through most of the arguments available.
The only variable not described is the NOEXIT argument since we created this for continuously running the JVM against the inbuilt test.
Being enabled by the user does not impact their usage as running a CryptoGuard scan against a single project will stop the JVM regardless of the flag.

\vspace*{15pt}\begin{figure}[H]
\begin{lstlisting}[language=make, caption=make, linerange={1-16, 56-58}]
Running the cryptoguard helper program
===============================================================================

Building a basic command
Please Note this does not verify whether the directory/files exist
Common abreviations:
	n = no
	y = yes
===============================================================================
JAR File flag: jar
APK File flag: apk
Directory of Source Code flag: source
Java File or Files flag: java
Class File or Files flag: class

What type of project by flag (jar/apk/source/java/class)? jar
===============================================================================
The build up command is:
java -jar cryptoguard -in jar -s test.jar -java /test/jdk8Path -in dependencies/ -main HelloWorld -m D -o results.json -new -vv -st -t -n -H -depth 10
\end{lstlisting}
	\caption{Excerpt from a successful CryptoSoule interactive build command.}
	\label{fig:CryptoSoule.py Sample One}
\end{figure}

Directly running the proper command, we show the user the start section (as shown in the figure \ref{fig:CryptoSoule.py Sample One}) and is immediately queried for their intentions.
After the user goes through and answers all the questions without any fail in validation, they will achieve a simple command to run on the command line (as shown in the last line of \ref{fig:CryptoSoule.py Sample One}).

\vspace*{15pt}\begin{figure}[H]
\begin{lstlisting}[language=make, caption=make, linerange={30-38}]
Output formats
Legacy flag: L extension: .txt
ScarfXML flag: SX extension: .xml
Default flag: D extension: .json

Would you like to specify the output format by flag (n/SX/L/D)? SX

Would you like to specify the output file (n/file)? test.json
Please enter a valid file for the output type.
\end{lstlisting}
	\caption{Excerpt from a failure example of the CryptoSoule interactive build command.}
	\label{fig:CryptoSoule.py Sample Two}
\end{figure}

This interactive command line runs through all the various input, requesting argument directly from the user and validating the cross dependencies between specifications.
Throughout this, there is validation to ensure file extensions match the argument and the source matches the arguments passed in.
As shown from figure \ref{fig:CryptoSoule.py Sample Two}, there is a validation error on the result file, as it should match the extension based on the flag.

\subsection{Argument Option Enumeration} \label{se:arg_enumeration}
The python script associated with the project also displays more helpful information about some of the more difficult arguments.
These arguments also determine various constraints upon other arguments as well, (cross-dependencies).
As an example, we display project types by name and as a flag for the argument.
The developers first view upon this may not be completely intuitive (as shown below in figure \ref{fig:Source Type Enum}).

\vspace*{15pt}\begin{figure}[H]
\begin{lstlisting}[language=bash, linerange=6-20]
	JAR File accepts a .jar
	APK File accepts a .apk
	Directory of Source Code accepts a dir
	Java File or Files accepts a .java
	Class File or Files accepts a .class
Running the cryptosoule helper program
===============================================================================

Can write the results as the following output types:
===============================================================================
	Legacy accepts a .txt file output type.
	ScarfXML accepts a .xml file output type.
	Default accepts a .json file output type.
Running the cryptosoule helper program
===============================================================================
\end{lstlisting}
	\caption{An exception showing the project type enumeration.}
	\label{fig:Source Type Enum}
\end{figure}

The previous following is only a single example of the total 3 types of enumeration commands.
Listed below from the python script are the following available enumeration help (from the excerpt \ref{fig:Type Enum Main}).

\vspace*{15pt}\begin{figure}[H]
\begin{lstlisting}[language=bash, linerange=12-14]
	./cryptosouple.py projectType: Displays some information about the project types available to scan.
	./cryptosouple.py outputType: Displays some information about the various output types available to write out as.
	./cryptosouple.py exceptionType: Displays information about the standardized exceptions.
\end{lstlisting}
	\caption{An exception showing the type enumeration.}
	\label{fig:Type Enum Main}
\end{figure}

\section{Live Notebook} \label{se:Notebook}

An exceedingly popular and upcoming live representation of data and programs, \href{https://jupyter.org/}{Jupyter Notebooks} \cite{ProjectJupyter} are a live and open source way to easily represent projects to users.
Used within various open source and research projects, Jupyter Notebooks (Notebooks) are browser-based programs that host sections of programs through different programming languages, documentation, or even video-based documentation.
Though Notebooks supports various programming languages through plugins, they are mostly associated with Python.
Developers create the Jupyter Notebooks to easily share sections of code online.
Jupyter Notebooks have sections of code that save their output states and provide an environment to reproduce the results.
Used in tandem with Notebooks is \href{https://mybinder.org}{Binder} \cite{BinderTeam}, a live Jupyter Notebook service.
Taking live GitHub repository URLs, Binder takes GitHub Repositories and recreates them into a live Jupyter Instance hosted via Docker.
With this unison, Binder hosts publicly available accessible Notebooks accessed by anyone solely using a web browser, lowering the barrier of access to a minimum.

Within CryptoGuard are several Notebooks that aide the testing effort run through CryptoSoule.
The main Notebook within the repository documents the testing process and creates several graphs, to create a quick representation for easy consumption.
Integrating into an extremely custom environment (with various JDKs used), the Binder instance displays the results and see the live documentation within CryptoSoule.
This enables potential consumers to quickly see information about CryptoGuard via a web browser.
Notably there is the Java Jupyter Notebook, \href{https://github.com/SpencerPark/IJava}{IJava} \cite{SpencerPark} that provides the Java programming language support for the Notebook.
Using Binder technology there is the potential to create test cases within the Notebook that directly interact with the CryptoGuard source code.
Due to the environment variables required, this is currently underway installing the Android SDK.

\section{Live Repository} \label{se:GitHubPackages}

With the intentions of distributing our program publicly and within the build tools of Java programs, we must ensure our programs are available to the build tools.
Instead of attempting to create an account onto one of the many popular build tool repositories, we instead choose to use the latest repository hosting site GitHub \cite{GitHubRepos}.
GitHub Packages support multiple build tools, not just Maven and Gradle for Java but also npm, Docker, NuGet, and RubyGems.
Providing an open platform for developers to link within the build tool, this provides an easier method for them to automatically download the latest project from an officially licensed source that is already publicly available.

\chapter{Discussion} \label{ch:discussion}
\section{Issues Traversed} \label{se:difficulties}

Despite the program milestones, there various issues arose.
There were many difficulties faced throughout the development of CryptoGuard.
CryptoGuards dependent libraries and original strict design created most of the issues.
These issues were ultimately solved but deserve specific mention.

\subsection{Soot Library Dependency} \label{sse:sootDependency}

One of the core libraries used within CryptoGuard is Soot, which is an immensely popular decompiler for Java projects.
Though the biggest change will be between Java 8 and Java 9 (due to the modularization of the JDK itself), there will still be depreciation and language support changes in the later version changes.
There is a vast number of guides for describing Soots standalone use or using the tool on the command line.
In contrast, some of the mentioned and highlighted projects making use of Soot are extending the functionality for use of data flow graphing or using it is capabilities to scan or interpret the source code flow (scanning code as an example).
Extending Soots library in this manner less extensively documented.
Given the first three options were already available within CryptoGuard, there seemed to be enough information to create the source and build file opportunities.
Though it is seemingly not stated anywhere in the documentation the ``front-end'' was not updated sufficiently to easily allow either source or build files, only archived distributable (such as Jar or APK files) or project directories.
Extending CryptoGuard to include working compiled Java Class files included experimenting with the existing architecture as well as the other posted projects (using Soot).
The Sable group actively maintains and responds to questions periodically about the Soot project.
By the time of this writing and the many growing changes between Java 8 and the latest versions of Java, there is a current lock on the Java version.
Though Soot is locked at a stable version on a Java version 6 versions below the latest seems staggering it is not a terrible issue.

\subsection{Streaming XML} \label{sse:StreamXML}

Described earlier within section \ref{sse:streaming}, our cloud consumer requested their output streamed directly to the file instead of building the object in memory and then writing the file out.
With the large volume per project they are using it makes sense, however given their XML based output ensuring the streamed file output validates against their schema provided thorough difficulty.
Not only ensuring the validation of the output but keeping it in line with the flat data structure ensured this problem would require a novel solution.
Though an advanced library, the marshalling output library \href{https://github.com/FasterXML/jackson}{FasterXML} only marshals on a class by class basis.

\vspace*{15pt}\begin{figure}[H]
	\begin{center}
		\includegraphics[scale=0.4]{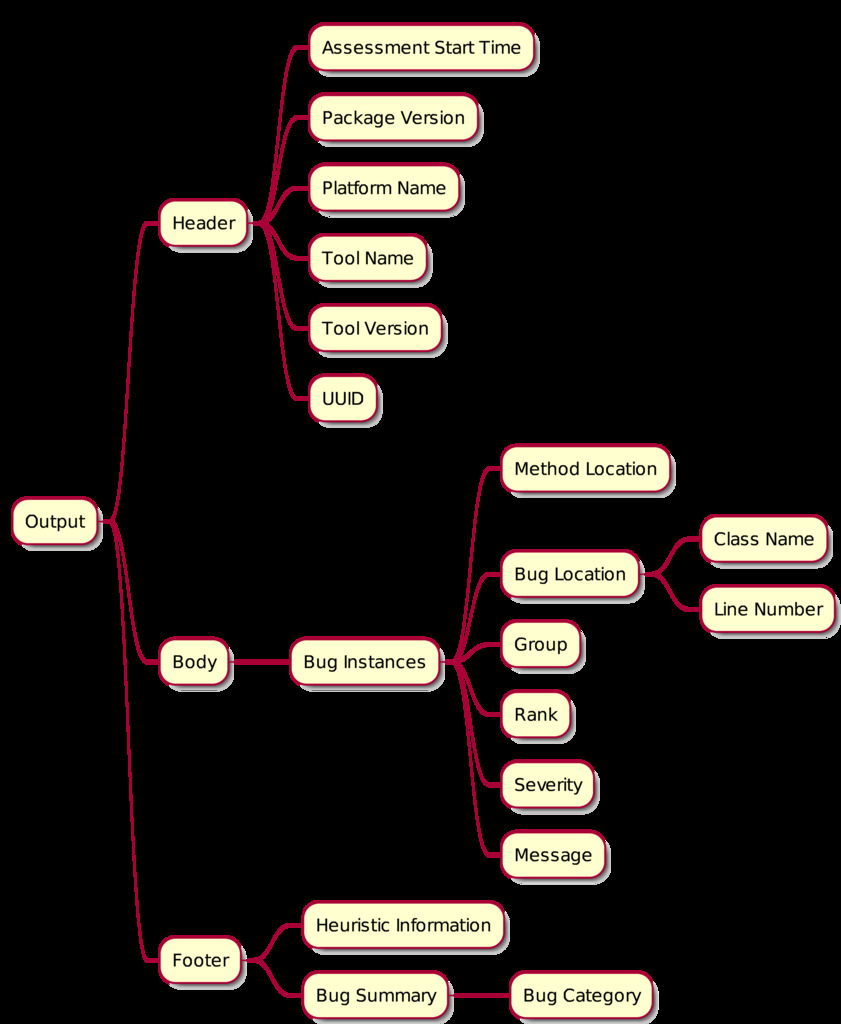}
	\end{center}
	\caption{The partition of the response document.}
	\label{fig:DocPartition}
\end{figure}

In order to ensure the flat output document structure could be passed both to streaming and non-streaming methodology, the output document had to be partitioned, as shown in figure \ref{fig:DocPartition} with some of the sample information contained per each section.
Representing the output within three different sections gives us enough room to allow customization to each representation.
The header section contains static information generated before the scanning begins.
The body section is the least strict section, in which it intentionally only handles the addition and manipulation of each cryptographic misuse (or Bug Instance).
We abstracted the body to both the streaming and non-streaming structure without the need to overlap in concerns.
The footer section contains the amalgamation of the information and will display any accumulated information, such as heuristic information or a summary on the cryptographic misuses.
Abstracting each section into two handlers per output structure, one streaming and one non-streaming.
This structural divide provides enough rationing of the document that each output structure can even choose to use the section or not.
For example, while the ScarfXML output structure uses the Header to display information such as the Tool Name (for both streaming and non-streaming), the Legacy output structure only uses the body and footer section of the output (also for both streaming and non-streaming).

\section{Future Work} \label{se:improvement}

\subsection{Java Version} \label{sse:JavaVersions}

Following up on describing some of the under pinning of CryptoGuard, the language versioning itself changed drastically.
Starting from 2016, Oracle has open-sourced the JDK to the OpenJDK community and the community has rapidly increased the pace of updates.
Though Oracle is not the sole maintainer of the JDK, they are still one (if not the) of the most active contributors to the JDK.
Previous under Oracle (and possibly Sun Microsystems) the JDK would take anywhere from 1 to 3 years to update a major version.
Since the OpenJDK community has taken over, it has decreased this time to 6 months for each major version.
Most of the major versions released are supported until the next release (or for 6 months) by the OpenJDK (different companies or communities may have different standards).
Along with Oracle every 3 years the next major version will be a Long-Term Support (LTS) version and supported for 3 following years.
Currently Java 8 and 11 are the current LTS where Java 8 support (from Oracle) will later this year.
At this point of writing Java 14 is the latest released JDK.
OpenJDK deprecated other non-LTS Java versions\footnote{Java Versions 9, 10, 12, and 13}.

Locked on Java 8 Version, CryptoGuard is still under the most popular Java Version used today.
When OpenJDK deprecates Java 8 we will follow other developers' mentality and this program will ``hop'' or upgrade LTS versions.
With this mentality, the program will upgrade to only the next LTS version.
Given there will only be a new LTS version every 3 years, upgrading from Java 8 is Java 11 (currently out) and after that Java 17 while skipping the subsequent Java versions.
Missing out on roughly 6 major Java versions, many developers do this to ensure the JDK they are programs target are supported for the next 3 years instead of the normal 6 months each non-LTS is supported by the community.

\subsection{Output Methodologies} \label{sse:output_methods}

With the removal of Java Cobra due to its removal from Java 9, FasterXML, the current library used in its place for marshalling the output provides extended and easy usage.
Already provided within this library are 10 \cite{FasterXML} different formats supporting various sub-domains, including plugin development, big data, or machine learning.
Instead of creating to a strict format out like the specified format for SWAMP (ScarfXML) and creating a generic format (like the default JSON format) would allow easier integration of CryptoGuard into any other platform.
Thus, extending CryptoGuard into other platforms as an API while keeping a loosely coupled relationship.

\subsection{Native Build Tool Validation} \label{sse:validation}

Listed within the CryptoGuard Binder is the main testing notebook that displays all the test results manually created.
The testing Notebook provides necessary test isolation and an easier medium to display test results.
Between the creation and destruction of Soot environments between test cases, there is residual memory within the JVM.
CryptoGuard tests override source files and memory potentially providing a ``bleeding effect'' ensuring false negatives.
Overall, this ensures Gradle tests break with false negatives executed natively.

Within the testing emphasis is a stricter approach to adding more thorough test cases.
Though there are around 60 tests, most of them are Integration Tests, passing the inputs directly into the entrance of the program this only lightly verifies the output.
Full test and memory isolation ensured forward progress and passing tests.
Adding more vigorous test cases, unit tests or method tests for example would help build stronger confidence and results within the community.

\subsection{Securing the IoT} \label{sse:IoTPlug}

\begin{figure}[H]
	\begin{center}
		\includegraphics[width=\textwidth]{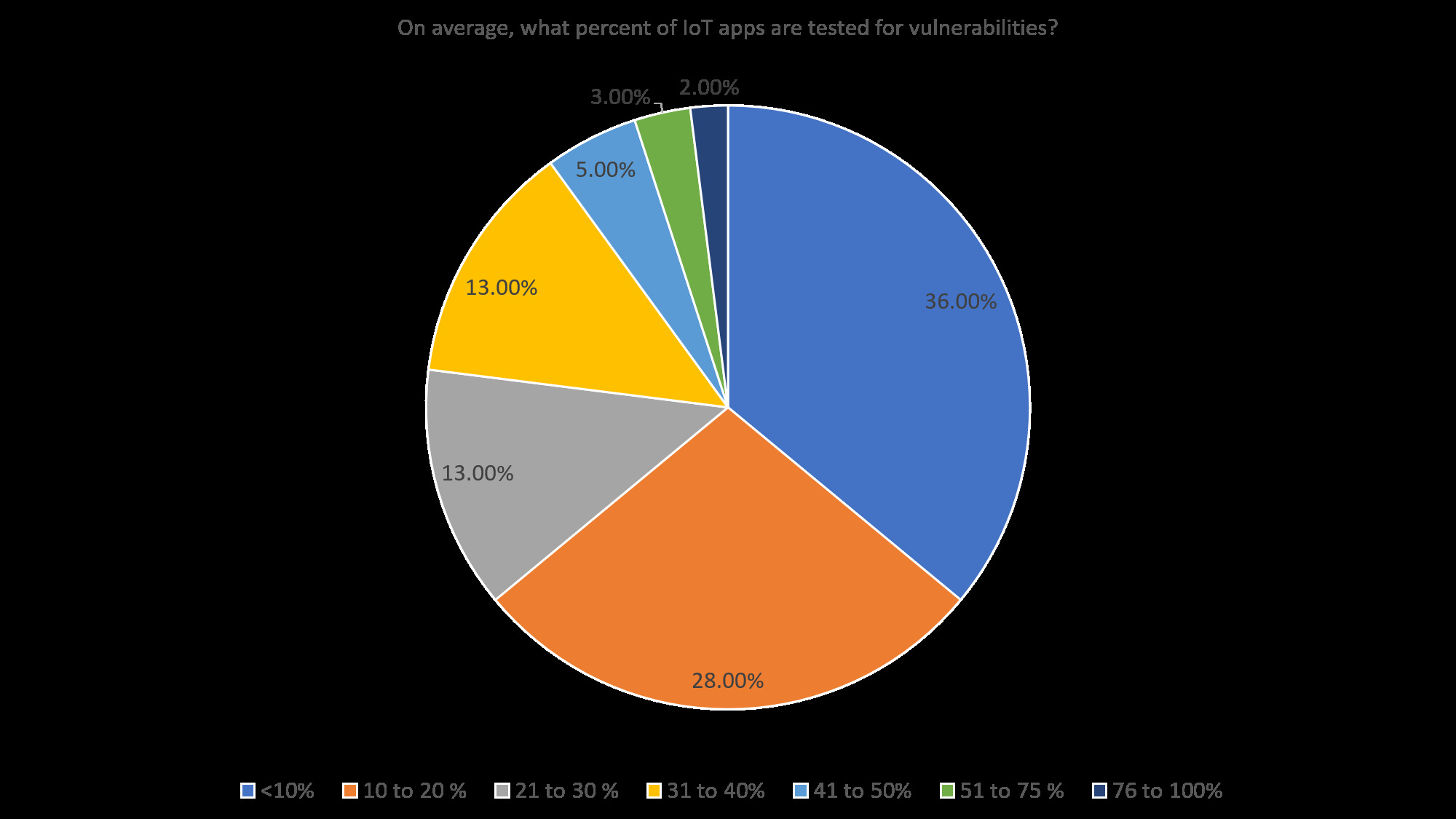}
	\end{center}
	\caption{Raw results depicting the overall testing effort of IoT applications \cite{PonemonInstituteLLC2017}.}
	\label{fig:Pono_AppTested}
\end{figure}

Securing software throughout the technology community not only deals with web apps but also through the IoT.
Though the IoT cannot handle the typical enterprise scale of programs it suffers from similar development and security issues.
Agile increased the speed of development \cite{Agile_People} including IoT applications.
A team studied a wide range of responses\footnote{593 of 16,450 initial participants} from a range of developers that at least 50 percent of their companies use between 51 to 500 IoT applications\cite{PonemonInstituteLLC2017}.
Although a majority\footnote{70 percent are somewhat to very concerned \cite{PonemonInstituteLLC2017} } of the companies are concerned with the potential of a hack through IoT applications, only about 10 percent of companies test up to half of the used IoT applications which depicted in figure\ref{fig:Pono_AppTested}.

\vspace*{15pt}

\begin{figure}[H]
	\begin{center}
		\includegraphics[width=\textwidth]{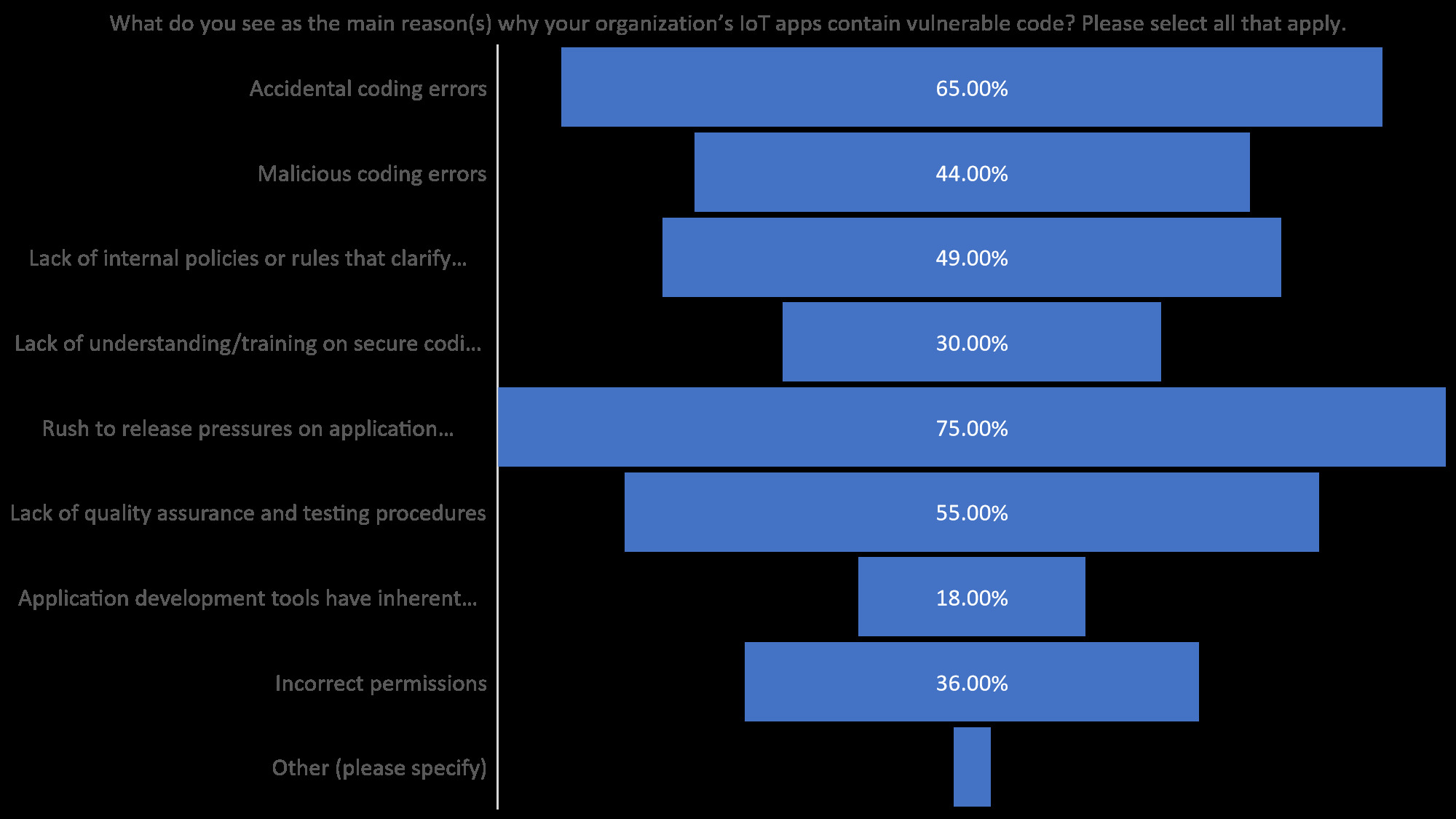}
	\end{center}
	\caption{Raw results depicting the perceived location of IoT vulnerabilities \cite{PonemonInstituteLLC2017}.}
	\label{fig:Pono_Where}
\end{figure}

Though it was determined that most developers believe securing IoT apps is very difficult\footnote{54\% rated 9 or 10 (Very difficult) \cite{PonemonInstituteLLC2017}}, the same developers also believe the issues lie within the development cycle itself.
Perceived as difficult to secure, deadlines introduce majority of the vulnerable code, accidental programming, or lack of QA or testing as depicted in figure \ref{fig:Pono_Where}.
Upon its design Java operates on various operating systems and architecture types.
Java is used within such low-power devices within the IoT.
As described within section \ref{se:Plugins}, ensuring tools are within build process ensures a seamless integration with the project.
Enhancing CryptoGuard to work within build tools, verifying it can catch majority of the issues introduced from figure \ref{fig:Pono_Where} could expand its potential and usage.

\vspace*{15pt}

\section{Conclusion} \label{se:Conclusion}

Throughout this work, we have expanded the capabilities of CryptoGuard triple fold.
Expanding its reach throughout the build systems, this allows a wider breadth of reach throughout the SDLC, at the lower abstraction within the developers' hands.
Though the Soot dependency locks CryptoGuard on an old JDK version, throughout surveys encompassing various mature developers from some of the big companies within the Java landscape confirm this is still the version used by clear majority.
Utilizing the online and free to use GitHub Package repository in combination with the Maven and Gradle plugins also allows developers to pull the latest packages from within the repository within their build system automatically.
Continuing the accession, continuing the push into the publicly and privately used system SWAMP will ensure CryptoGuard's use in the cloud.

\nocite{*}
\bibliographystyle{ieeetr}
\bibliography{Sections/references}

\begin{thebibliography}{10}

\bibitem{Agile_Cycle}
L.~Williams and A.~Cockburn, ``{Agile software development: it's about feedback
  and change},'' {\em Computer}, vol.~36, no.~6, pp.~39--43, 2003.

\bibitem{Koebler2020a}
J.~Koebler, ``{An ‘Off-the-Shelf, Skeleton Project': Experts Analyze the App
  That Broke Iowa - VICE}.''
  \url{https://www.vice.com/en\_us/article/3a8ajj/an-off-the-shelf-skeleton-project-experts-analyze-the-app-that-broke-iowa},
  2020.
\newblock Accessed March 9, 2020.

\bibitem{Rahaman2019}
S.~Rahaman, Y.~Xiao, S.~Afrose, F.~Shaon, K.~Tian, M.~Frantz, M.~Kantarcioglu,
  and D.~Yao, ``{Cryptoguard: High precision detection of cryptographic
  vulnerabilities in massive-sized Java projects},'' {\em Proceedings of the
  ACM Conference on Computer and Communications Security}, pp.~2455--2472,
  2019.

\bibitem{Agile_People}
A.~Cockburn and J.~Highsmith, ``{Agile software development, the people
  factor},'' {\em Computer}, vol.~34, no.~11, pp.~131--133, 2001.

\bibitem{PonemonInstituteLLC2017}
{Ponemon Institute LLC}, ``{2017 Study on Mobile and Internet of Things
  Application Security},'' {\em Ponemon Institute}, no.~January, p.~30, 2017.

\bibitem{Franceschi-Bicchierai2017}
L.~Franceschi-Bicchierai, ``{Equifax Was Warned - VICE}.''
  \url{https://www.vice.com/en\_us/article/ne3bv7/equifax-breach-social-security-numbers-researcher-warning},
  2017.
\newblock Accessed March 9, 2020.

\bibitem{Newman2017}
L.~Newman, ``{The Equifax Breach Was Entirely Preventable | WIRED}.''
  \url{https://www.wired.com/story/equifax-breach-no-excuse/}, 2017.
\newblock Accessed March 9, 2020.

\bibitem{Koebler2020}
J.~Koebler and E.~Maiberg, ``{Here's the Shadow Inc. App That Failed in Iowa
  Last Night - VICE}.''
  \url{https://www.vice.com/en\_us/article/y3m33x/heres-the-shadow-inc-app-that-failed-in-iowa-last-night},
  2020.
\newblock Accessed March 9, 2020.

\bibitem{Meraki2019}
C.~Meraki and Proprivacy, {\em {USENIX Supporters USENIX Patrons Bloomberg •
  Facebook • Google Microsoft • NetApp USENIX Benefactors Amazon • Oracle
  Two Sigma • VMware USENIX Partners USENIX Security '19 Sponsors Platinum
  Sponsor Diamond Sponsor Silver Sponsors Bronze Sponsors Gold Spon}}.
\newblock 2019.

\bibitem{10.5555/2486788.2486877}
B.~Johnson, Y.~Song, E.~Murphy-Hill, and R.~Bowdidge, ``{Why Don't Software
  Developers Use Static Analysis Tools to Find Bugs?},'' in {\em Proceedings of
  the 2013 International Conference on Software Engineering}, ICSE '13,
  pp.~672--681, IEEE Press, 2013.

\bibitem{OWASP}
OWASP, ``{OWASP Foundation, the Open Source Foundation for Application
  Security}.'' \url{https://owasp.org/}.
\newblock Accessed February 25, 2020.

\bibitem{OWASPTool}
OWASP, ``{Source Code Analysis Tools | OWASP}.''
  \url{https://owasp.org/www-community/Source\_Code\_Analysis\_Tools}, 2020.
\newblock Accessed February 25, 2020.

\bibitem{Rahaman2019a}
S.~Rahaman, Y.~Xiao, S.~Afrose, K.~Tian, M.~Frantz, N.~Meng, B.~P. Miller,
  F.~Shaon, M.~Kantarcioglu, and D.~Yao, ``{Poster: Deployment-quality and
  accessible solutions for cryptography code development},'' in {\em
  Proceedings of the ACM Conference on Computer and Communications Security},
  pp.~2545--2547, Association for Computing Machinery, nov 2019.

\bibitem{Oraclea}
Oracle, ``{History of Java Technology}.''
  \url{https://www.oracle.com/technetwork/java/javase/overview/javahistory-index-198355.html}.
\newblock Accessed February 11, 2020.

\bibitem{Oracle}
Oracle, ``{Java SE Name and Version Changes}.''
  \url{https://www.oracle.com/technetwork/java/javase/namechange-140185.html}.
\newblock Accessed March 11, 2020.

\bibitem{Javatpoint}
Javatpoint, ``{Learn Maven Tutorial - javatpoint}.''
  \url{https://www.javatpoint.com/maven-tutorial}.
\newblock Accessed March 19, 2020.

\bibitem{Apache}
Apache, ``{Maven – Welcome to Apache Maven}.''

\bibitem{GradleHome}
Gradle, ``{Gradle Build Tool}.'' \url{https://gradle.org/}.
\newblock Accessed March 19, 2020.

\bibitem{Gradle2020}
Gradle, ``{Gradle | Releases}.'' \url{https://gradle.org/releases/}, 2020.
\newblock Accessed March 13, 2020.

\bibitem{LightbendRelease}
Lightbend, ``{Release 1.0.0 {\textperiodcentered} sbt/sbt {\textperiodcentered}
  GitHub}.'' \url{https://github.com/sbt/sbt/releases/tag/v1.0.0}.
\newblock Accessed March 19, 2020.

\bibitem{Lightbend}
Lightbend, ``{sbt - The interactive build tool}.''
  \url{https://scala-sbt.org/}.
\newblock Accessed March 19, 2020.

\bibitem{GoogleBazelRelease}
Google, ``{Bazel 1.0 - Bazel}.''
  \url{https://blog.bazel.build/2019/10/10/bazel-1.0.html}.
\newblock Accessed March 19, 2020.

\bibitem{GoogleBazel}
Google, ``{Bazel - a fast, scalable, multi-language and extensible build system
  - Bazel}.'' \url{https://bazel.build/}.
\newblock Accessed March 19, 2020.

\bibitem{SWAMP}
SWAMP, ``{Software Assurance Marketplace}.''
  \url{https://continuousassurance.org/}.
\newblock Accessed February 27, 2020.

\bibitem{DHS2011}
DHS, ``{Past Solicitations | DHS BAA}.''
  \url{https://baa2.st.dhs.gov/portal/public/processRequest?eurl=AAAAAAEytBoAAAFuYT8BtQAUQUVTL0NCQy9QS0NTNVBhZGRpbmcAgAAQABAAAQIDBAUGBwgJCgsMDQ4PAAAAcMUP8ssYOu8SxeEfopmq%2F3JxVJpIrI76XzFWBZiTRiyCifJD0j16VQoZINHrrHUF17cm56S8jVIZTN53bDdu1EqawWuNYoQ0fRI6glx650CTJlGBpp5yA8w8O3cvKT3S5V1b2JVMunL3HAwHAvH7Em0AFK6vJ7ezs3d2Pd4VUrdAdjpXvQ0L},
  2011.
\newblock Accessed March 25, 2020.

\bibitem{MIRSwamp}
ContinuousAssurance, ``{Software Assurance Marketplace}.''
  \url{https://www.mir-swamp.org/}.
\newblock Accessed March 25, 2020.

\bibitem{McGillUniversity/SableResearchGroup}
McGillUniversity/SableResearchGroup, ``{What is Soot? | Soot}.''
  \url{https://sable.github.io/soot/}.
\newblock Accessed March 18, 2020.

\bibitem{GithubFrantz2019}
M.~Frantz, ``{Issue loading Java Class {\textperiodcentered} Issue {\#}1140
  {\textperiodcentered} Sable/soot}.''
  \url{https://github.com/Sable/soot/issues/1140}, 2019.
\newblock Accessed March 26, 2020.

\bibitem{ContinuousAssuranced}
ContinuousAssurance, ``{resultparser/scarf\_v1.2.xsd at master
  {\textperiodcentered} mirswamp/resultparser {\textperiodcentered} GitHub}.''
  \url{https://github.com/mirswamp/resultparser/blob/master/xsd/scarf\_v1.2.xsd}.
\newblock Accessed March 9, 2020.

\bibitem{Tutorialspoint_OO}
Tutorialspoint, ``{Design Pattern - Overview - Tutorialspoint}.''
  \url{https://www.tutorialspoint.com/design\_pattern/design\_pattern\_overview.htm}.
\newblock Accessed March 25, 2020.

\bibitem{ObjectOrientedDesign}
ObjectOrientedDesign, ``{Design Patterns | Object Oriented Design}.''
  \url{https://www.oodesign.com/}.
\newblock Accessed March 25, 2020.

\bibitem{BlackWasp2009}
BlackWasp, ``{Gang of Four Design Patterns}.''
  \url{http://www.blackwasp.co.uk/GofPatterns.aspx}, 2009.
\newblock Accessed March 25, 2020.

\bibitem{SpringFrameworkGuru}
SpringFrameworkGuru, ``{Gang of Four Design Patterns - Spring Framework
  Guru}.'' \url{https://springframework.guru/gang-of-four-design-patterns/}.
\newblock Accessed March 25, 2020.

\bibitem{Oracle2020}
Oracle, ``{Oracle Java SE Support Roadmap}.''
  \url{https://www.oracle.com/java/technologies/java-se-support-roadmap.html},
  2020.
\newblock Accessed March 11, 2020.

\bibitem{Javatpoint2018}
Javatpoint, ``{Java Versions | Java Version History - Javatpoint}.''
  \url{https://www.javatpoint.com/java-versions}, 2018.
\newblock Accessed March 11, 2020.

\bibitem{OpenJDK2010}
OpenJDK, ``{OpenJDK FAQ},'' 2010.

\bibitem{Oracle2020_Roadmap}
Oracle, ``{Oracle Java SE Support Roadmap}.''
  \url{https://www.oracle.com/java/technologies/java-se-support-roadmap.html},
  2020.
\newblock Accessed March 11, 2020.

\bibitem{Spectrum/IEEE2016}
Spectrum/IEEE, ``{Interactive: The Top Programming Languages 2016 - IEEE
  Spectrum}.''
  \url{https://spectrum.ieee.org/static/interactive-the-top-programming-languages-2016},
  2016.
\newblock Accessed March 11, 2020.

\bibitem{Spectrum/IEEE2018}
Spectrum/IEEE, ``{Interactive: The Top Programming Languages 2018 - IEEE
  Spectrum}.''
  \url{https://spectrum.ieee.org/static/interactive-the-top-programming-languages-2018},
  2018.
\newblock Accessed March 11, 2020.

\bibitem{Vermeer2020}
Vermeer, ``{JVM Ecosystem Report 2020 | Snyk}.''
  \url{https://snyk.io/blog/jvm-ecosystem-report-2020/}, 2020.
\newblock Accessed March 9, 2020.

\bibitem{Maple2018}
S.~Maple and A.~Binstock, ``{JVM Ecosystem Report 2018 | Snyk}.''
  \url{https://snyk.io/blog/jvm-ecosystem-report-2018/}, 2018.
\newblock Accessed March 9, 2020.

\bibitem{Apache2019}
Apache, ``{Maven – Maven Releases History}.''
  \url{https://maven.apache.org/docs/history.html}, 2019.
\newblock Accessed March 13, 2020.

\bibitem{EclipseFoundation}
EclipseFoundation, ``{EE4J FAQ | The Eclipse Foundation}.''
  \url{https://www.eclipse.org/ee4j/faq.php}.
\newblock Accessed March 13, 2020.

\bibitem{EclipseFoundation2018}
EclipseFoundation, ``{Jakarta EE Community Survey of 1,800+ Java Developers
  Reveals “Cloud Native” Top Requirement in Platform's Evolution}.''
  \url{https://www.globenewswire.com/news-release/2018/04/24/1485972/0/en/Jakarta-EE-Community-Survey-of-1-800-Java-Developers-Reveals-Cloud-Native-Top-Requirement-in-Platform-s-Evolution.html},
  2018.
\newblock Accessed March 11, 2020.

\bibitem{Milinkovich2019}
EclipseFoundation, ``{The Cloud Native Imperative — Results from the 2019
  Jakarta EE Developer Survey | Life at Eclipse}.''
  \url{https://eclipse-foundation.blog/2019/05/10/results-2019-jakarta-ee-developer-survey/},
  2019.
\newblock Accessed March 11, 2020.

\bibitem{JetBrains}
Jetbrains, ``{JetBrains: Developer Tools for Professionals and Teams}.''
  \url{https://www.jetbrains.com/}.
\newblock Accessed March 13, 2020.

\bibitem{Spectrum/IEEE2019}
Spectrum/IEEE, ``{Interactive: The Top Programming Languages 2019 - IEEE
  Spectrum}.''
  \url{https://spectrum.ieee.org/static/interactive-the-top-programming-languages-2019},
  2019.
\newblock Accessed March 11, 2020.

\bibitem{Chumak2017}
A.~Chumak, ``{Developer Ecosystem Survey: Raw Data Are Here | JetBrains
  Blog}.''
  \url{https://blog.jetbrains.com/blog/2017/09/11/developer-ecosystem-survey-raw-data-are-here/},
  2017.
\newblock Accessed February 25, 2020.

\bibitem{Chumak2018}
A.~Chumak, ``{Developer Ecosystem Survey 2018: Raw Data Available | JetBrains
  Blog}.''
  \url{https://blog.jetbrains.com/blog/2018/11/05/developer-ecosystem-survey-2018-raw-data-available/},
  2018.
\newblock Accessed February 25, 2020.

\bibitem{Chumak2019}
A.~Chumak, ``{Developer Ecosystem Survey 2019: Raw Data Available | JetBrains
  Blog}.''
  \url{https://blog.jetbrains.com/blog/2019/08/13/developer-ecosystem-survey-2019-raw-data-available/},
  2019.
\newblock Accessed February 25, 2020.

\bibitem{OpenJDK}
OpenJDK, ``{JDK}.'' \url{https://openjdk.java.net/projects/jdk/}.

\bibitem{meng2018secure}
N.~Meng, S.~Nagy, D.~Yao, W.~Zhuang, and G.~A. Argoty, ``{Secure coding
  practices in java: Challenges and vulnerabilities},'' in {\em Proceedings of
  the 40th International Conference on Software Engineering}, pp.~372--383,
  2018.

\bibitem{ProjectJupyter}
ProjectJupyter, ``{Project Jupyter | Home}.'' \url{https://jupyter.org/}.
\newblock Accessed March 27, 2020.

\bibitem{BinderTeam}
BinderTeam, ``{Binder}.'' \url{https://mybinder.org/}.
\newblock Accessed March 27, 2020.

\bibitem{SpencerPark}
SpencerPark, ``{GitHub - SpencerPark/IJava: A Jupyter kernel for executing Java
  code.}.'' \url{https://github.com/SpencerPark/IJava}.
\newblock Accessed March 27, 2020.

\bibitem{GitHubRepos}
GitHub, ``{GitHub Packages: Your packages, at home with their code
  {\textperiodcentered} GitHub}.'' \url{https://github.com/features/packages}.
\newblock Accessed March 27, 2020.

\bibitem{FasterXML}
FasterXML, ``{GitHub - FasterXML/jackson: Main Portal page for the Jackson
  project}.'' \url{https://github.com/FasterXML/jackson}.
\newblock Accessed March 27, 2020.

\end{thebibliography}

\appendix

\begin{appendices}
	\chapter{Diagrams} \label{app:diagrams}
	\section{CryptoGuard Internal Structure} \label{ase:CryptoGuard}
	\vfill
	\begin{figure}[H]
		\begin{center}
			\includegraphics[width=\linewidth]{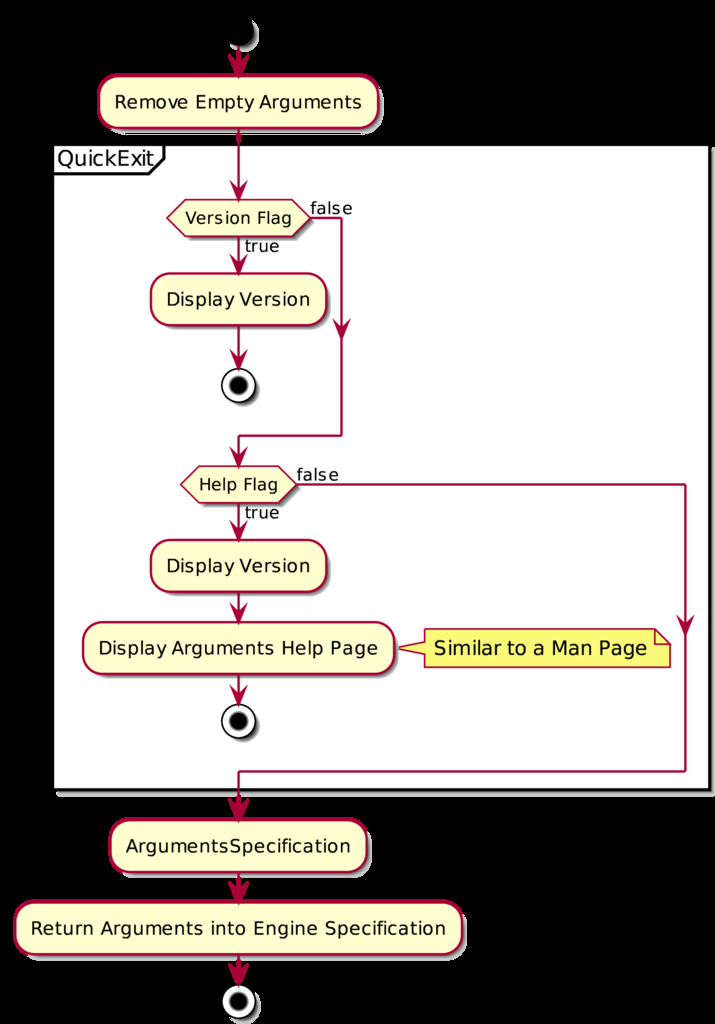}
		\end{center}
		\caption{The first stage of CryptoGuard, examining the flags and trimming the empty arguments.}
		\label{fig:cryptoguard_entry}
	\end{figure}
	\vfill
	\newpage
	\vfill
	\begin{figure}[H]
		\begin{center}
			\includegraphics[width=\linewidth]{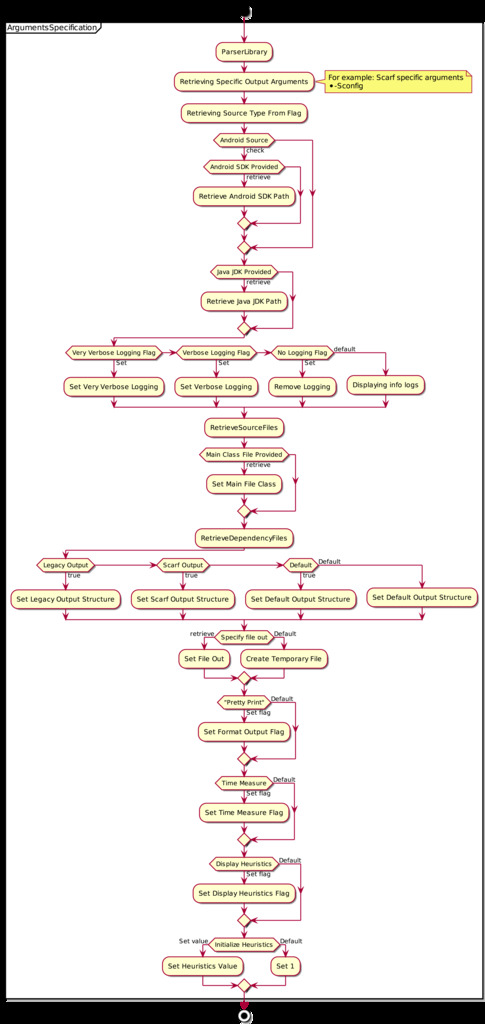}
		\end{center}
		\caption{The revised argument parsing, checking the generalized flags and files.}
		\label{fig:cryptoguard_argspec}
	\end{figure}
	\vfill
	\newpage
	\vfill
	\begin{figure}[H]
		\begin{center}
			\includegraphics[width=\linewidth]{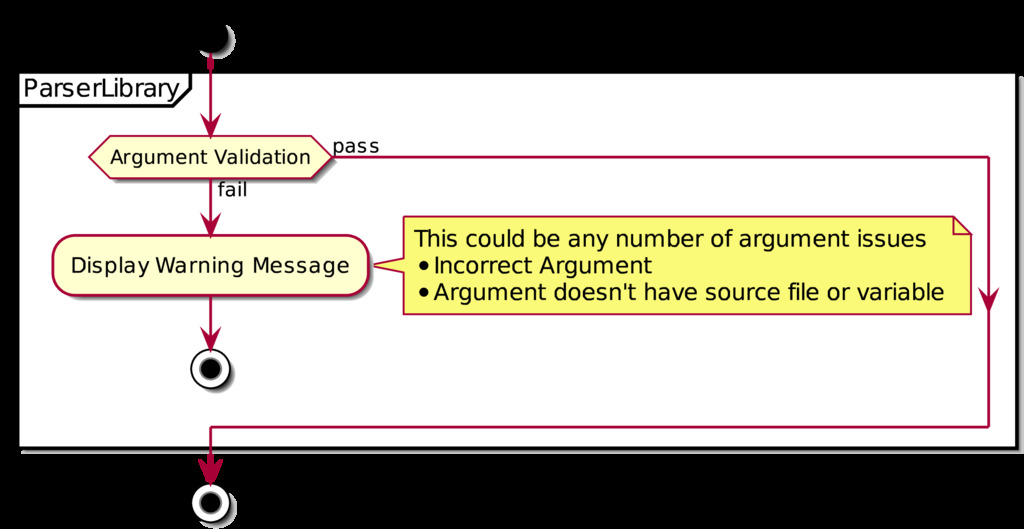}
		\end{center}
		\caption{The argument parser library, verifying the arguments and handling the error message.}
		\label{fig:cryptoguard_parserlib}
	\end{figure}
	\vfill
	\newpage
	\vfill
	\begin{figure}[H]
		\begin{center}
			\includegraphics[width=\linewidth]{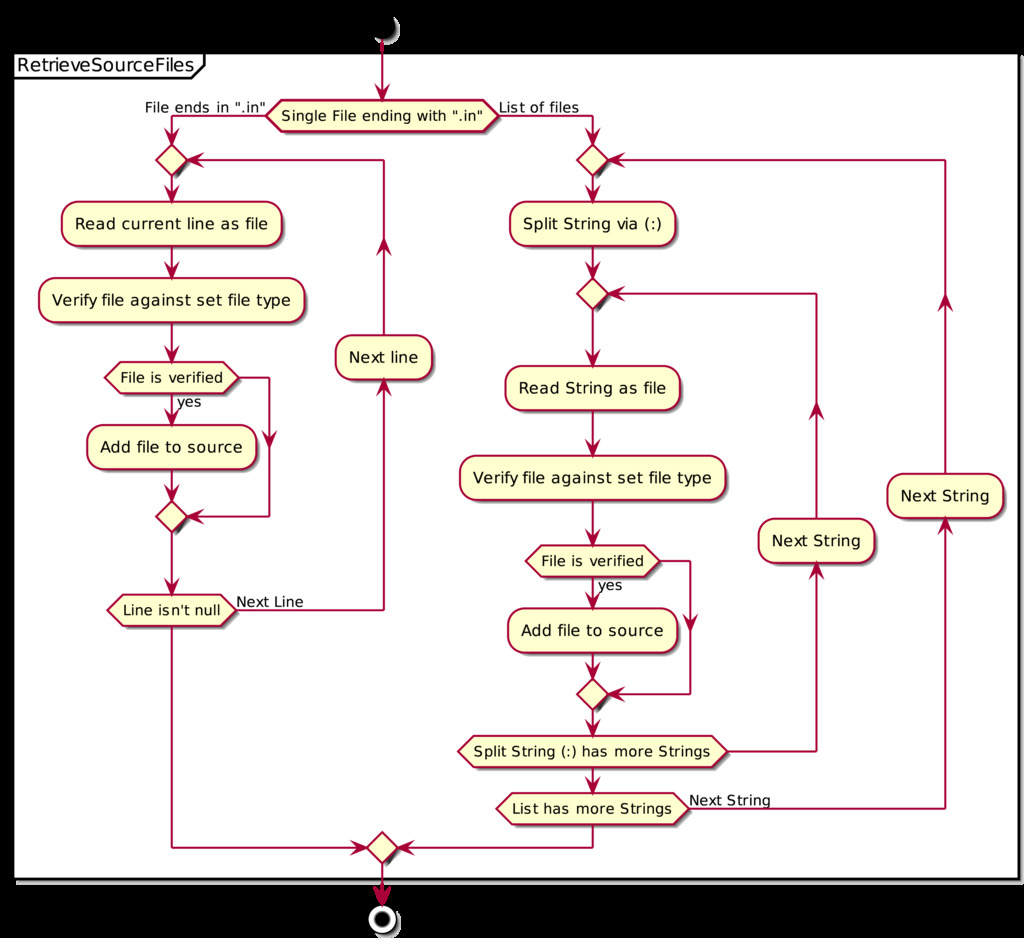}
		\end{center}
		\caption{The method to retrieve the source files based on the three different ways provided.}
		\label{fig:cryptoguard_sourcedeps}
	\end{figure}
	\vfill
	\newpage
	\vfill
	\begin{figure}[H]
		\begin{center}
			\includegraphics[width=\linewidth]{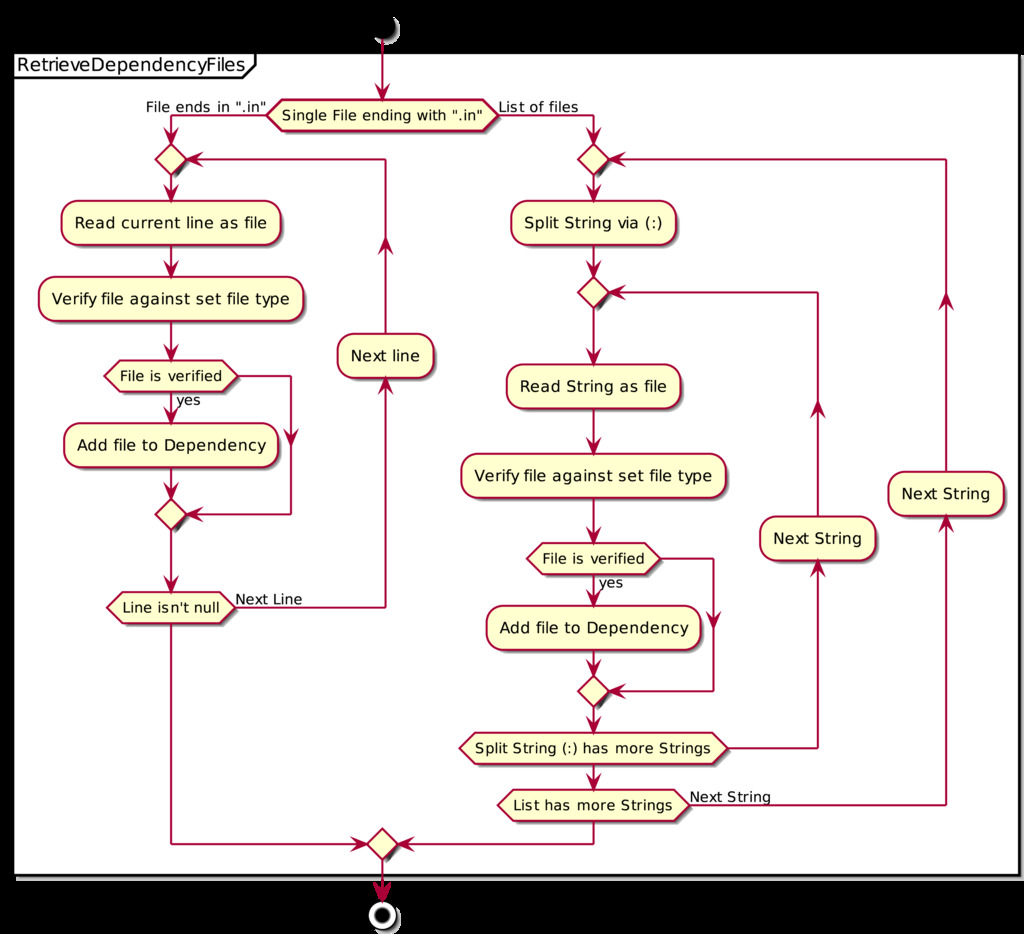}
		\end{center}
		\caption{The method to retrieve the dependency files based on the three different ways provided.}
		\label{fig:cryptoguard_retrievedeps}
	\end{figure}
	\vfill
	\newpage
	\vfill
	\begin{figure}[H]
		\begin{center}
			\includegraphics[width=\linewidth]{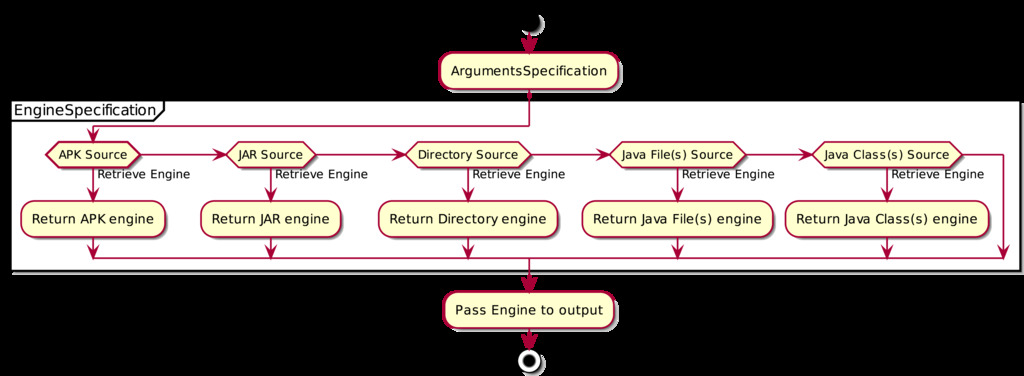}
		\end{center}
		\caption{The dynamically retrieved engine method, determining the scanning method.}
		\label{fig:cryptoguard_enginespec}
	\end{figure}
	\vfill
	\newpage
	\vfill
	\begin{figure}[H]
		\begin{center}
			\includegraphics[width=\linewidth]{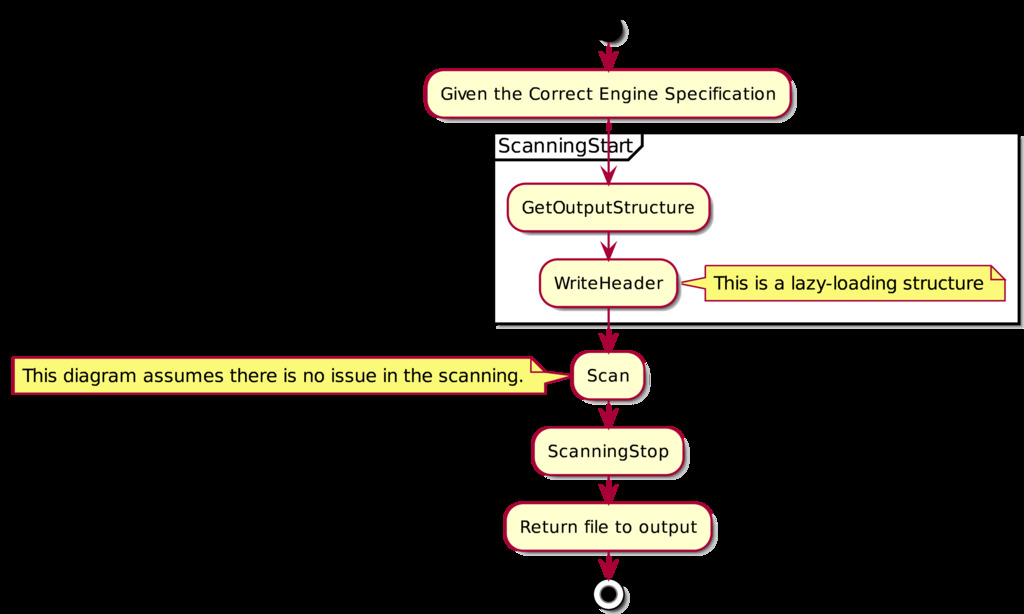}
		\end{center}
		\caption{The scanning overhead of CryptoGuard.}
		\label{fig:cryptoguard_scanning overhead}
	\end{figure}
	\vfill
	\newpage
	\vfill
	\begin{figure}[H]
		\begin{center}
			\includegraphics[width=\linewidth]{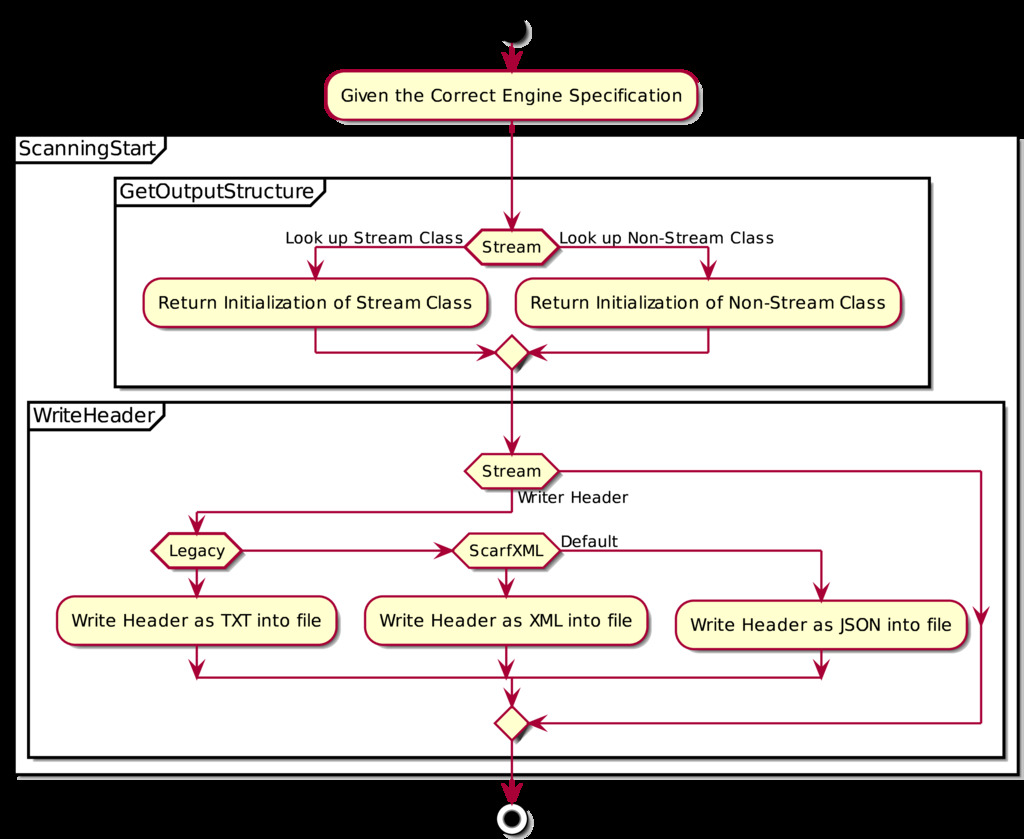}
		\end{center}
		\caption{The method for the start scanning section.}
		\label{fig:cryptoguard_start_scanning}
	\end{figure}
	\vfill
	\newpage
	\vfill
	\begin{figure}[H]
		\begin{center}
			\includegraphics[width=\linewidth]{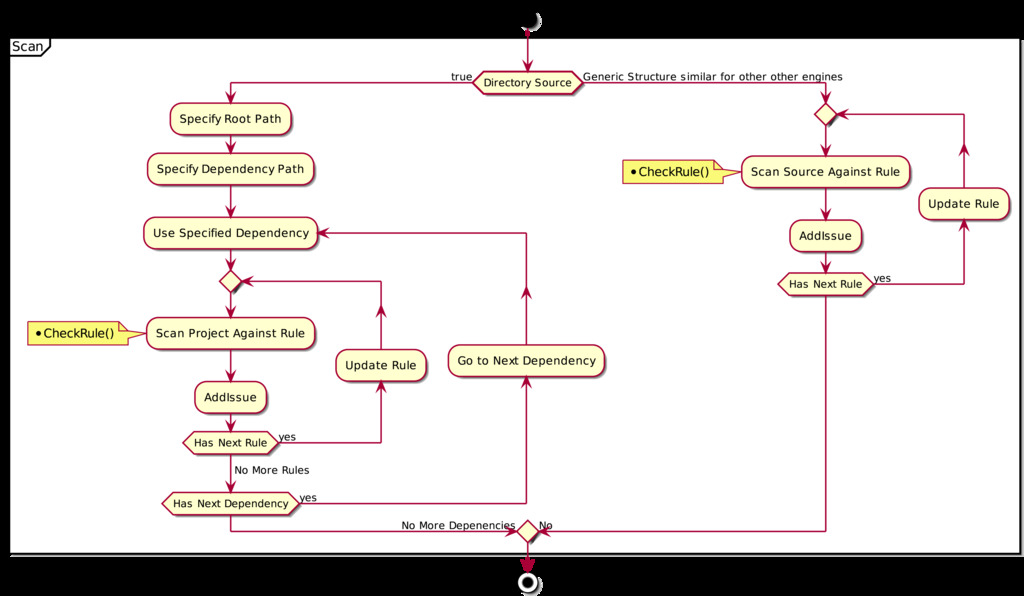}
		\end{center}
		\caption{The overview of the scanning method.}
		\label{fig:cryptoguard_overview_scanning}
	\end{figure}
	\vfill
	\newpage
	\vfill
	\begin{figure}[H]
		\begin{center}
			\includegraphics[width=\linewidth]{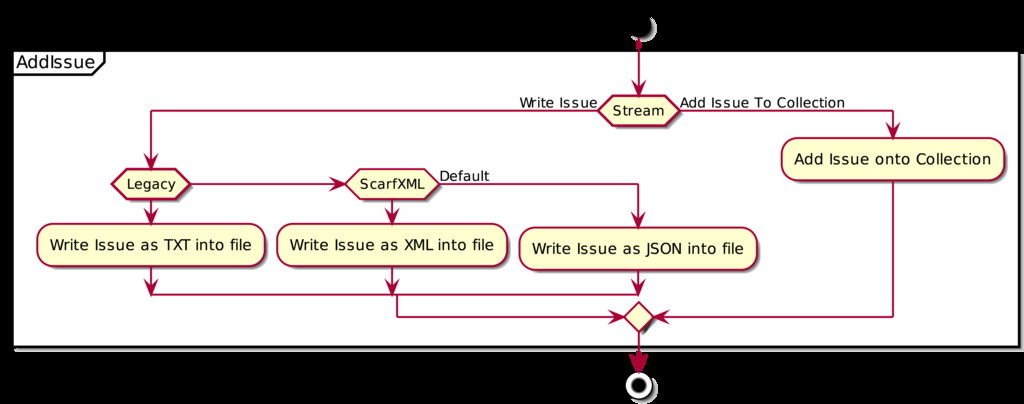}
		\end{center}
		\caption{The add issue method explored.}
		\label{fig:cryptoguard_add_issue}
	\end{figure}
	\vfill
	\newpage
	\vfill
	\begin{figure}[H]
		\begin{center}
			\includegraphics[width=\linewidth]{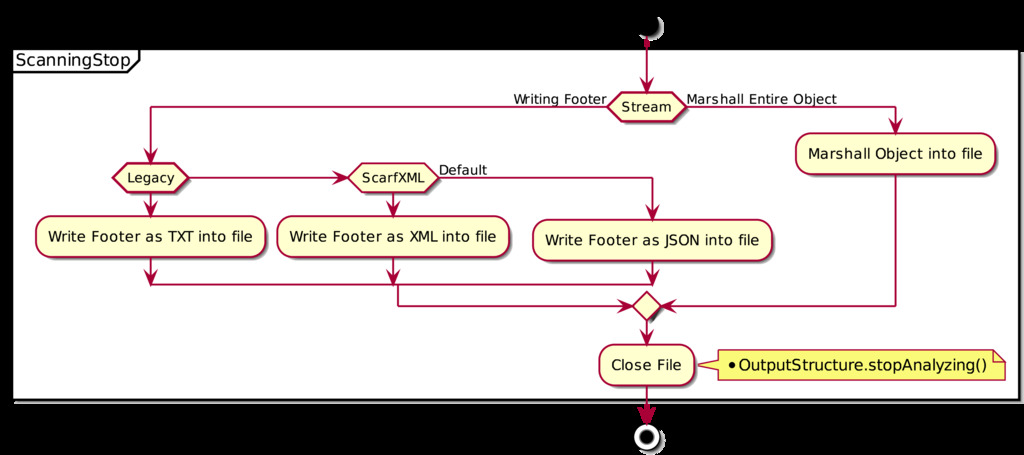}
		\end{center}
		\caption{The overall flow of the stop scanning method.}
		\label{fig:cryptoguard_scanner_overhead}
	\end{figure}
	\vfill
	\newpage
	\vfill
	\begin{figure}[H]
		\begin{center}
			\includegraphics[scale=0.75]{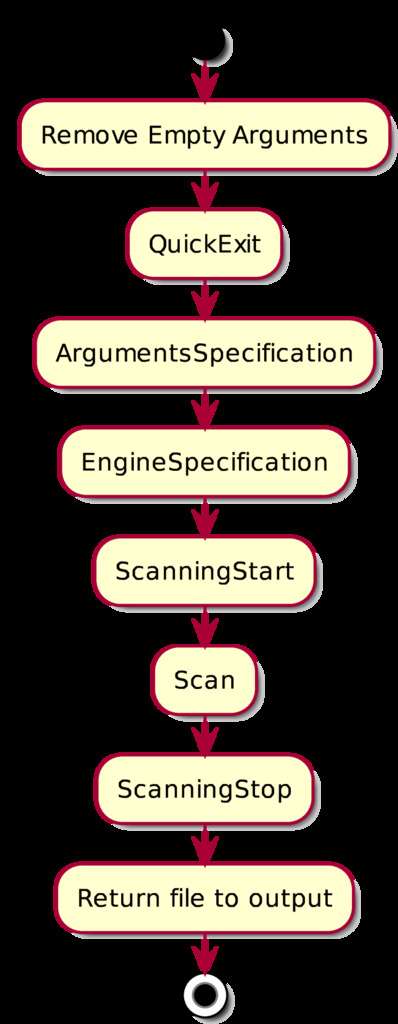}
		\end{center}
		\caption{The overview of the overall CryptoGuard flow.}
		\label{fig:cryptoguard_light_overhead}
	\end{figure}
	\vfill
	\newpage
	\vfill
	\section{Plugin Structures} \label{ase:Plugins}
	\begin{figure}[H]
		\begin{center}
			\includegraphics[width=\linewidth]{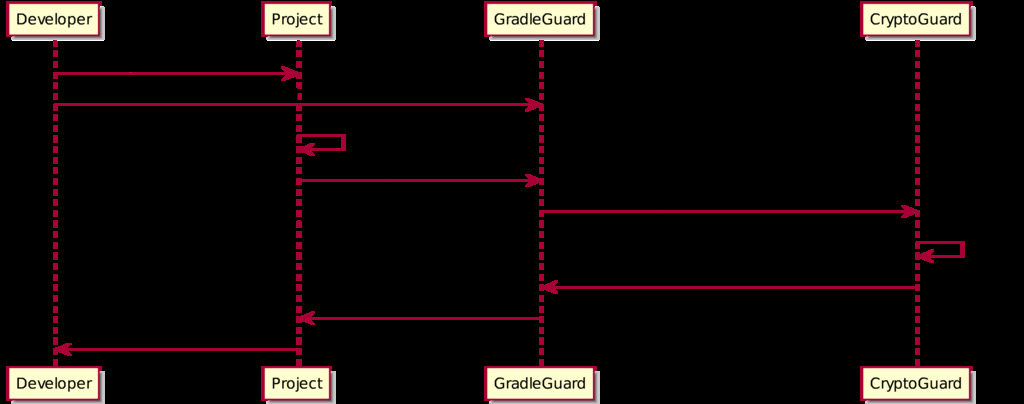}
		\end{center}
		\caption{The usage flow for the GradleGuard plugin.}
		\label{fig:plugin_gradleguard}
	\end{figure}
	\vfill
	\newpage
	\vfill
	\begin{figure}[H]
		\begin{center}
			\includegraphics[width=\linewidth]{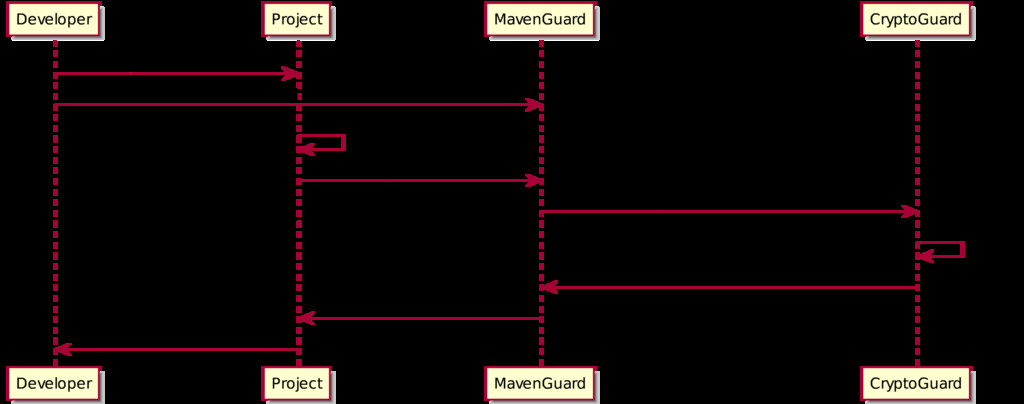}
		\end{center}
		\caption{The usage flow for the MavenGuard plugin.}
		\label{fig:plugin_mavenguard}
	\end{figure}
	\vfill
	\newpage
	\vfill
	\section{Live Notebook Environment} \label{ase:NoteBook Env}
	\begin{figure}[H]
		\begin{center}
			\includegraphics[width=\linewidth]{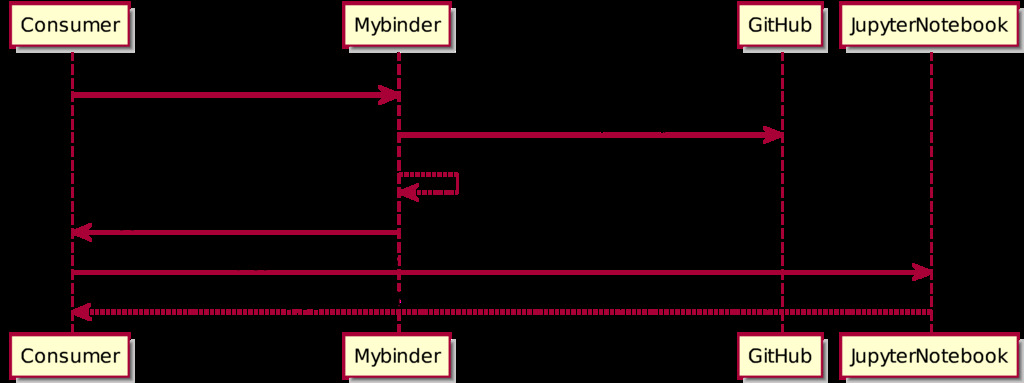}
		\end{center}
		\caption{The usage flow for the Mybinder and Jupyter Notebook flow.}
		\label{fig:NoteBook_Activity}
	\end{figure}
\end{appendices}

\end{document}